# Electron-beam-driven collective self-hybridized exciton polaritons

Parsa Darman[1ǂ]*, Maximilian Black[1 ǂ]*, Prabhdeep Singh[1], Victor DeManuel-Gonzales[1], Sara Darbari[2], Masoud Taleb[1,3], Fatemeh Chahshouri[1,3], Nahid Talebi[1,3]*

1 Institute of Experimental and Applied Physics, Kiel University, 24418 Kiel, Germany

2 Faculty of Electrical and Computer Engineering, Tarbiat Modares University, Tehran 1411713116, Iran

3 Kiel Nano, Surface, and Interface Science KiNSIS, Kiel University, 24118 Kiel, Germany

ǂ These authors contributed equally to this work.

Email: talebi@physik.uni-kiel.de, darman@physik.uni-kiel.de, black@physik.uni-kiel.de

## Abstract:

Self-hybridized exciton polaritons in layered semiconductors emerge from the hybridization of excitonic resonances with confined photonic modes in films and provide a versatile platform for controlling light–matter interactions at the nanoscale. Yet, the nature of exciton-photon interactions and collective and synchronous strong-coupling of multiple excitons to the same photonic mode are largely unexplored. Here, we use complementary momentum-resolved photoluminescence and cathodoluminescence spectroscopy to compare self-hybridized exciton polaritons in Ruddlesden–Popper perovskite flakes under optical and electron-beam excitation. We find that cathodoluminescence exhibits a remarkably larger polaritonic level repulsion than photoluminescence, revealing an enhanced effective coupling strength under electron-beam excitation. We attribute this enhancement to the collective coherent excitation of multiple excitons by fast electrons, which couple to the same photonic mode and increase the interaction strength according to a $\sqrt{N}$ scaling, where up to $N = 23$ is revealed. By reducing the electron kinetic energy, we further uncover a crossover to a partially incoherent excitation regime in which higher-order Fabry–Pérot exciton-polariton resonances disappear, leaving only lowest-polariton branches and yet, coherent transition radiation. This transition arises because low-energy electrons excite emitters over extended spatial volumes, washing out the phase coherence required to sustain a collective coupling to form polaritonic resonances. Our work shows that electron beams can actively modify collective light–matter coupling in layered semiconductors and establishes cathodoluminescence as a route to access excitation regimes beyond all-optical excitation schemes.

## Introduction:

Strong coupling between excitonic and photonic resonances gives rise to hybrid light–matter quasiparticles named exciton-polaritons. The theoretical foundation of exciton–photon hybridization was established by Hopfield[1], while semiconductor microcavity experiments subsequently demonstrated the characteristic anti-crossing behavior in the strong-coupling regime[2]. Strong coupling was later realized down to the single-emitter limit in semiconductor quantum-dot cavities[3-5], followed by studies of external control and quantum nonlinearities in strongly coupled quantum-dot–cavity systems[6-8 7–10]. When multiple equivalent emitters interact coherently with the same electromagnetic mode, the interaction can additionally become collective. The interaction of a single two-level emitter with a quantized cavity mode is described by the Jaynes–Cummings model[9], while its extension to *N* emitters provided by Tavis and Cummings predicts a collectively enhanced coupling strength, $g_n = \sqrt{N} g_1$, for *N* identical oscillators coupled coherently to a common mode[10]. Such collective enhancement has been demonstrated experimentally in coupled emitter-cavity systems such as superconducting circuits and semiconductor nanocavities containing multiple emitters[11,12].

A prominent factor which underpins the strength of light-matter interaction is the oscillator strength of the excitonic excitation. Two-dimensional Ruddlesden–Popper perovskites (RPPs) are particularly suitable for exciton–polariton physics because their naturally stacked quantum-well structure gives rise to an enhanced oscillator strength, and a binding energy of several hundred millielectronvolts at room temperature[13-15] . For $(BA)_2PbI_4$ RPP, for example, an exciton binding energy of approximately 260 meV has been reported[16]. Room-temperature strong exciton–photon coupling in layered perovskites was demonstrated in early microcavity experiments[17,18], followed by further demonstrations of coherently coupled and nonlinear polariton states[19,20]. Even without an external cavity, a semiconductor thin film may sustain confined optical modes, which strongly interact with excitons. These so-called self-hybridized exciton polaritons were first observed in thin layers of semiconducting transition metal dichalcogenides[21,22]. Two classes of self-hybridized exciton polaritons coexist, namely, Fabry–Pérot and guided-wave modes that spatially overlap with the excitonic medium. The dispersion line for the former is positioned within the light cone, whereas for the latter extends from the vacuum light line to the light line inside the material[23]. Extending the research to perovskites, self-hybridized polaritonic emission was demonstrated in layered perovskites[24], while subsequent studies revealed multimode polariton formation, energy transfer between polariton branches, lasing, and long-range propagation[25-27]. Self-hybridization has since been extended as well to very large and ultrastrong coupling strengths in layered excitonic materials[28].

Next to all-optical spectroscopy techniques, cathodoluminescence (CL) spectroscopy has established itself as a prominent method for exploring light-matter interactions at the nanoscale[29-35]. CL has been used extensively as well to study the photophysics of excitons in semiconductors[36-38]; however, incoherent CL emission has been long regarded as the dominant mechanism for semiconductors rather than coherent exciton energy-transfer mechanisms. This is mainly due to the fact that particularly for semiconductors, cascaded excitations of electron-hole pairs by the secondary or back-scattered electrons and their out-of-phase relaxation lead to an incoherent burst of photon bunches[39-43]. This picture, nevertheless, is substantially altered within the strong-coupling regime, where photon-mediated couplings between emitters lead to a synchronous excitation of excitons and their coupling to the photonic modes. In this realm, self-hybridized exciton polaritons in thin transmission metal dichalcogenide have been explored using CL spectroscopy[44-47]. However, the collective aspects of such resonances have been not yet explored.

Here, we extend the previous studies on the exploration of self-hybridized exciton-polaritons with CL spectroscopy to the regime of collective excitation of multiple excitons and their strong coupling to the same photonic mode, within the context of Tavis-Cummings model. Benefiting from the strong oscillator strength of exciton excitations in thin perovskite flakes excited by the electron beam in a laterally confined volume smaller than the wavelength of the photonic mode, we demonstrate a substantially enhanced strong-coupling strength, when compared to photoluminescence (PL) studies. We demonstrate that this enhancement in the coupling strength is due to an increased effective number of excitonic oscillators participating collectively in the electron-driven interaction, consistent with the $\sqrt{N}$ scaling of collective light–matter coupling. By varying the electron kinetic energy, we further demonstrate a crossover from multiple Fabry–Pérot exciton–polariton resonances to a regime dominated by the transition radiation from the upper surface and an incoherent excitation. Our results therefore identify the collective excitation pathway as an additional degree of freedom for controlling self-hybridized exciton-polaritons and show that fast electrons can access collective coupling regimes not easily addressable by optical excitation.

## Results and discussion

The investigated sample consists of mechanically exfoliated RPP flakes placed on a 40-nm-thick gold film deposited on a glass substrate. Considering that in electron microscopy the substrate has to be conductive to avoid charging, the vacuum-RPP-gold system proves to be a suitable system to investigate polariton formation in CL and simultaneously for correlative PL explorations. For further information about the synthesis and sample preparation, see the Methods section.

A moving electron inside an electron microscope carries an evanescent near-field, prominently acting as a localized, ultra-broadband, and background-free electromagnetic source for exciting the material excitations, and couples to the photonic local density of states of the sample under the exploration[41,48-50]. Therefore, when interacting with the RPP film, the photonic modes are excited, such as the Fabry–Pérot resonances, which are hosted by the cavity formed by the thin crystal itself, with its surfaces acting as semi-transparent mirrors (Fig. 1a). Accordingly, the photons become localized within the RPP layer, enabling a long interaction time with the excitons and the subsequent formation of exciton-polaritons. The moving electron also excites transition radiation, which can be understood in terms of the dynamic dipole formed by the electron and its induced image charge in the film. As the electron crosses the interface, the abrupt change of this dynamic dipole gives rise to broadband radiation with a characteristic dipole-like far-field radiation pattern[51].

When additionally multiple excitonic states are excited coherently, they form a superposition, which synchronously couple to the same photonic mode. The coherent coupling of *N* excitonic oscillators gives rise to a collective bright state with an enhanced oscillator strength, resulting in a collectively enhanced coupling strength $g = \sqrt{N} g_1$, where $g_1$ is the coupling strength of a single excitonic oscillator to the considered cavity mode.

The interaction mechanisms of an electron beam with the RPP film depends as well on the electron's initial kinetic energy. An electron beam with a high kinetic energy exchanges energy with the RPP layer, but keeps most of its energy upon exiting the material, implying a relatively narrow interaction volume and a short interaction time. This fact will be substantiated further below by the Monte-Carlo simulations, mapping the trajectory and the kinetic energies of backscattered and secondary electrons for electron beams at different initial kinetic energies. Under this condition, the excitons are excited in an ultrafast manner within this narrow interaction volume as well, which is deep sub-wavelength compared to the wavelength of the cavity photons, therefore forming a synchronous excitation scheme.

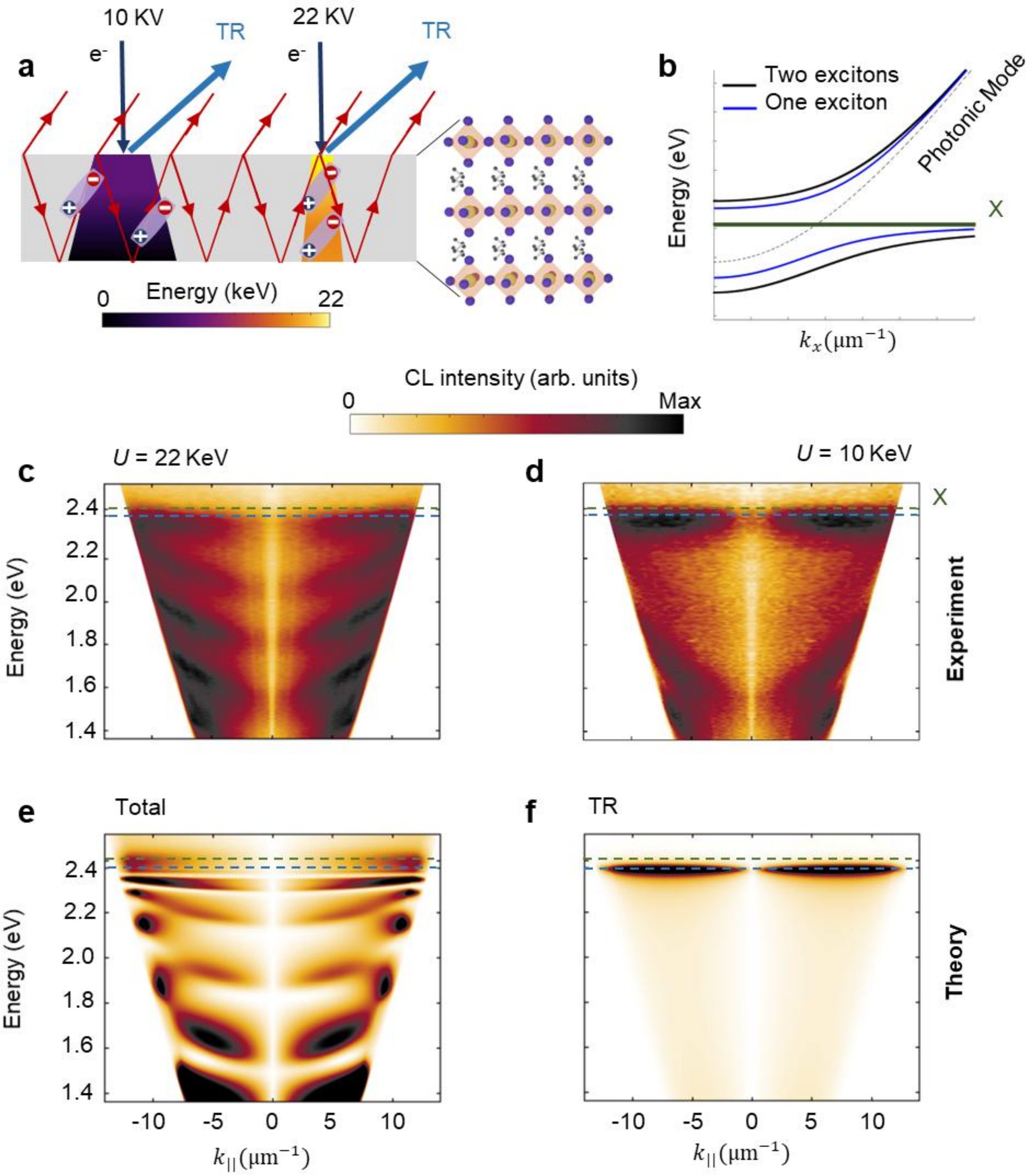

**Figure 1. Optical response of a thin RPP layer under electron beam irradiation. (a)** Cathodoluminescence excitation scheme under varying electron acceleration, producing both transition radiation and Fabry–Pérot modes, with the latter coupling to simultaneously excited excitons. (**b**) Energy-momentum dispersion relation of a photonic mode coupling to an exciton resonance, specifically demonstrating an enhanced level repulsion for a higher number of coupled excitonic states. Momentum-resolved cathodoluminescence spectral map of a 680 nm thick RPP layer for exciting the structure with (c)an electron beam with the kinetic energy of 22 keV, where the response is dominated by polaritonic modes, and (**d**) an electron beam with a kinetic energy of 10 keV, dominated by the transition radiation. Calculated momentum resolved cathodoluminescence maps for (e) total radiation and (f) only transition radiation.

Contrary to this, an electron beam with a lower kinetic energy loses almost all of its energy within an RPP film with the thickness of 680 nm and hardly exits the material at its bottom layer, leading to an extended interaction volume. Therefore, the excitons are excited out-of-phase with respect to the evanescent field of the moving electron beam over this extended volume and within a longer interaction time. Thus, the special and temporal distribution of energy loss of the electron beam and excited secondary or backscattered electrons dictate the properties of the exciton-photon coupling. For an electron beam with 22 keV kinetic energy, numerous excitons are excited within a short interaction time synchronously and in phase with respect to the evanescent field of the moving electron, which respectively lead to a synchronization and a defined phase correlation between excitons themselves. Hence, a larger acceleration voltage and electron velocity effectively leads to a coherent in-phase coupling of multiple excitons to one photonic mode. This fact is further substantiated by the angle-resolved CL spectra, demonstrating the formation of Fabry–Pérot modes for an electron at the initial kinetic energy of $U = 22\,\mathrm{keV}$, as compared to the CL map acquired for an electron at the initial energy of $U = 10\,\mathrm{keV}$ (Fig. 1c and d), as will be further elaborated in more details below.

First, we shortly discuss the theory of exciton-polariton formation as understood by the Hopfield model (See Supplementary Note S1 for more details).[1,23] The interactions between the photonic and excitonic wave functions results in an avoided crossing of the dispersion at the exciton resonance and effectively creates two polaritonic energy branches, as shown schematically in Fig. 1b. Generally, the branches above the exciton energy are called upper polariton branches (UPBs), while their counterparts below the exciton energy are referred to as lower polariton branches (LPBs). The dispersions in Fig. 1b were calculated using the Hopfield model, considering one photonic cavity mode and one, respectively two excitonic oscillators coupled to the same photonic mode. The effect of collective coupling to one photonic mode is visualized here, as the coupling of two excitons with one cavity mode increases the level repulsion between the polaritonic branches. The quantity governing this repulsion is the coupling strength $g$, and it is equal to the splitting of the Hopfield energy levels under tuned excitation ($\omega_c\left(k_{\|}\right) = \omega_e$), as

$$E_H = \frac{1}{2}\hbar\left(\omega_c\left(k_{\|}\right) + \omega_e\right) \pm \frac{1}{2}\hbar\sqrt{\left(\omega_c\left(k_{\|}\right) - \omega_e\right)^2 + 4g^2} \qquad (1)$$

where $\omega_c\left(k_{\|}\right)$ and $\omega_e$ denote the frequencies of the cavity photon and of the exciton, respectively, and $k_{\|}$ is the wavenumber of the photons parallel to the surface of the RPP film (along the propagation direction of the polaritons). The Fabry–Pérot resonances are strongly dispersive and follow a parabola-like dispersion (dashed line in Fig. 1b), whereas the excitonic oscillator is dispersion-free (exciton diffusion is neglected). The exciton resonances in the bulk RPP happen at exactly $\hbar\omega_e = 2.43\,\mathrm{eV}$, as will be further resolved below using PL and absorption spectroscopy. Increasing the number of excitonic basis effectively increases the level repulsion. For the case of *N* excitonic oscillators interacting in phase with a single photonic mode, the coupling factor is altered to $\sqrt{N}g$, as understood by the Tavis Cummings model (see Supplementary Note S1). This has been shown in the case of a given number of atomic oscillators in a cavity[52], as well as for a known number of layered quantum wells[53].

The cathodoluminescence measurements with the initial kinetic energies of 22 keV and 10 keV for the electron beams showcase the fundamental difference of the CL response to fast and slow electrons (see

Fig. 1c and d). With the 22 keV electron beam traversing a flake with the thickness of 860 nm, LPBs were observed in the form of six distinguished branches in the momentum resolved energy maps. Particularly, the flattening of the parabola-like dispersions near to the excitonic resonance and the avoided crossing is unambiguously reproduced in all CL maps acquired from RPP flakes, and is the signature of exciton-polariton formation. However, the UPBs could not be resolved due to strong damping arising from absorption in $(BA)_2PbI_4$. The absorption cross section of the RPP increases substantially at photon energies above the excitonic resonance, consistent with the larger imaginary part of the permittivity in this spectral range (Supplementary Fig. 1a). The absence of clearly resolved upper polariton branches in experimentally measured dispersion maps has also been reported for other perovskite polaritonic systems[19,20].

When using an electron beam with an initial kinetic energy of 10 keV, the CL intensities show that the polaritonic resonances are suppressed, and a uniform dispersion-less radiation peak at 2.4 eV becomes dominant (see Fig. 1d), which we identify as the transition radiation, as substantiated theoretically below.

To better understand the different radiation mechanisms of the electron beam, we analytically calculate the momentum-resolved CL spectra using the no-recoil approximation, where the electron beam is modelled as a coherent current distribution inducing a polarization within the sample (see Supplementary Note S2 for the full derivation)[49,50]. The momentum-resolved CL spectrum ($\Gamma^{\mathrm{CL}}$) is modelled as the radiated power to the far-field, and is captured as

$$\Gamma^{\mathrm{CL}}\left(\omega,k_{\|}\right)=\Gamma^{\mathrm{FP}}\left(\omega,k_{\|}\right)+\Gamma^{\mathrm{TR}}\left(\omega,k_{\|}\right)+\Gamma^{\mathrm{INT}}\left(\omega,k_{\|}\right), \tag{2}$$

where $\Gamma^{\mathrm{FP}}$, $\Gamma^{\mathrm{TR}}$, and $\Gamma^{\mathrm{INT}}$ are the contributions of the Fabry–Pérot resonances, transition radiation from the top surface, and the interference between them to the overall CL spectrum. For an electron at the kinetic energy of 22 keV traversing an RPP flake with the thickness of 860 nm, six LPBs are clearly reproduced in the analytical momentum-resolved CL maps (Fig. 1e), in close analogy with the experimental data (Fig. 1c). More interestingly, the contribution of the transition radiation to the overall CL map demonstrates a clear peak at 2.4 eV, and an ultra-broadband feature over the entire energy ranges below 2.4 eV (Fig. 1f). In fact, the transition radiation is captured as

$$\Gamma^{\mathrm{TR}}\left(\omega,k_{\|}\right)=\frac{e^2k_0^4}{4\pi^2}\left|\frac{\left(\varepsilon_r\left(\omega\right)-1\right)}{\left(\varepsilon_r\left(\omega\right)k_0^2-k_{\|}^2-\left(\dfrac{\omega}{v_e}\right)^2\right)\left(k_0^2-k_{\|}^2-\left(\dfrac{\omega}{v_e}\right)^2\right)}\right|^2\frac{k_{\|}^2}{\omega\varepsilon_0}\mathrm{Re}\left\{\sqrt{k_0^2-k_{\|}^2}\right\}, \tag{3}$$

where $k_0=\omega c^{-1}$ is the wavenumber of light in the vacuum, $e$ is the elementary charge, $\varepsilon_r\left(\omega\right)$ is the dispersive relative permittivity for RPP (see Supplementary Fig. S3a), and $v_e$ is the electron velocity. The transition-radiation peak from the top surface is observed at photon energies where the real part of the permittivity reaches a maximum, corresponding to an enhanced amplitude of the image charge induced by the moving electron. This behavior is also captured experimentally for slow electrons, when the dominant mechanism of the radiation turns to be the transition radiation and not the coherent Fabry-Pérot resonances. The spectral features of this radiation are only related to the RPP permittivity and not the flake thickness. Moreover, whereas the transition radiation intensity is dependent on the electron velocity, the latter does not lead to significant changes in the spectral distribution of the transition radiation.

Another competing mechanism is particularly the contribution of the guided-wave resonances with their dispersion lying outside the light cone (see the modal representation in Supplementary Fig. S4). For the asymmetric system of RPP films positioned on a reflecting gold substrate, the onset of the excitation of the guided-wave resonances at the light line in vacuum is displaced compared to the Fabry–Pérot resonances, resulting in isolated energy-momentum features near to the light line in the calculated momentum-resolved CL spectra (Fig. 1e, Supplementary Fig. S3 and 4).

**Self-hybridized polaritons explored by optical spectroscopy**

To discuss the influence of the RPP thickness and the substrate, we have performed PL spectroscopy with a 480 nm laser excitation and compared the results with optical absorption spectroscopy, for which a broadband light source is used (see the Methods section). These experiments have been performed for a thin RPP flake on glass as well as for a thick flake on gold-coated glass, and the results were closely compared. The PL spectrum of the 50 nm-thin flake shows a single peak at $\lambda = 510\,\mathrm{nm}$ ($E = 2.43\,\mathrm{eV}$) that coincides with the absorption peak (Fig. 2a), while no other resonances are observed. Since the Fabry–Pérot resonances of RPP layers with thicknesses below 120 nm are not resonant with the exciton energy, strong exciton–photon coupling does not occur. Consequently, the PL and absorption peaks originate from the direct emission and absorption of the excitons, respectively.

In contrast, for a 760 nm-thick RPP flake positioned on the gold-coated substrate, no spectral peak is observed at the exciton energy in either PL or absorption spectroscopy. Instead, a multitude of longer-wavelength resonances are revealed, whose resonance wavelengths match well between the PL and absorption spectra. These peaks emerge due to exciton–polariton formation and are identified as LPBs. Furthermore, their narrow spectral shape indicates a high cavity quality, which is attributed to the presence of gold as a highly reflective substrate.

**Density of coherently excited excitons affecting the coupling strength:**

CL and PL responses of the RPP film positioned on a gold-coated glass substrate are substantially different. This behavior is omnipresent among all RPP flakes we have studied and is not related to local morphologies and imperfections. To quantify this difference, we showcase the angle-resolved CL, reflection and PL spectroscopy performed subsequently on the same RPP flake with 808 nm thickness, and marked the position of the electron beam impact in the SEM image in Fig. 3a, as well as the location of the optical measurements in the reflection image in Figure 3b. The electron beam has a kinetic energy of 20 keV in these measurements. The laser spot and especially the electron beam can cause changes in the morphology of the flake, which appear as dark spots in the images. The radiation damage caused by the electron excitation though alters the flake quality only locally. Therefore, under careful acquisition time control, laser field intensity, and electron-beam current, we observe no change in the dispersion of the Fabry–Perot resonances while performing the measurements.

The flake thickness was determined by confocal optical profilometry using a MarSurf confocal microscope (MarSurf CM Explorer). The height profile in Figure 3c shows two elevated bubbles at electron beam impact positions, indicating a local thickening or lifting of the RPP layer. LPBs of the layered system appear as peaks in CL and PL and as dips in the reflectivity, with their dispersion exhibiting flattening and polaritonic anti-crossing behaviors near the exciton energy (Fig. 3d to f). In the optical measurements, the same dispersion for the LPBs is observed, whereas in the CL measurement, the modes are considerably flatter and appear to be redshifted. Notably, two types of modes with distinguished dispersions are observed in the CL maps, which is due to the coexistence of Fabry–Pérot resonances within the light cone and guided-wave polaritons, where the signatures of the latter are obvious only near the vacuum light line. An efficient peak-finding algorithm has been used to find the position of both resonances (see Supplementary Note S1), and the dashed lines in Fig. 3d depicts only the position of Fabry–Pérot exciton polaritons.

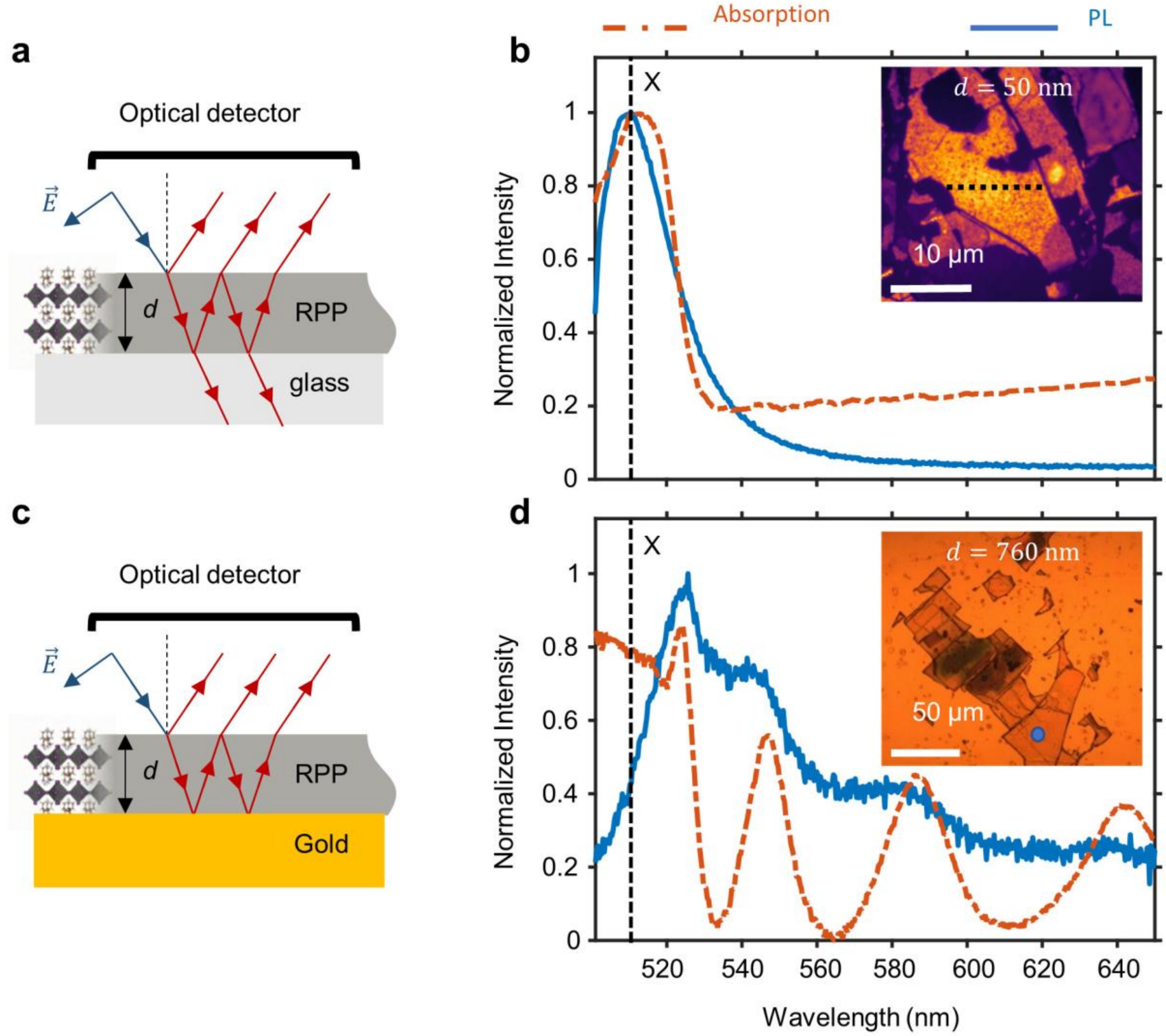

**Figure 2: Exciton luminescence for thin films and polariton formation for thick films. (a**) Schematic of the optical excitation of the air-RPP-glass open cavity system. (**b**) Absorption and PL spectra for a 50 nm thin RPP flake on glass, showing only the excitonic emission due to a lack of Fabry–Pérot resonances coupling to the exciton, and (**c**) Optical excitation scheme of the air-RPP-gold heterostructure. (**d**) Absorption and PL spectra for and RPP film with 760 nm thickness positioned on gold-coated glass substrate, where multiple Fabry–Pérot LPBs are observed.

No obvious shift of the polaritonic modes is observed with varying laser intensity in PL (Supplementary Fig. S5). This leads to the conclusion that, in the all-optical measurements presented here, the effective number of excitonic resonances coherently coupled to a given cavity photon mode remains unchanged. Although increasing the laser intensity results in a larger density of excited excitons within the focal region of the laser on the sample, no enhanced collective response is observed. This can be attributed to the relatively large excitation volume, with lateral dimensions comparable to the wavelength of light, such that the excitons experience different phases of the optical field. Consequently, an in-phase collective coupling of the excitons to the photonic mode is suppressed.

To quantify the level repulsion and therefore the coupling strength, we fit the polaritonic dispersions with momentum-resolved energies derived from the Hopfield model (see Supplementary Note S1). This resulted in varying coupling strengths, where the Fabry–Pérot modes closest to the exciton energy experienced the highest coupling strength, which is $g_{\mathrm{opt}} = 74\ \mathrm{meV}$. The variation of the coupling strength with mode number originates from the spatial modal distribution and (standing wave patterns of the Fabry–Pérot within the RPP film) and thereby the electric field distribution of cavity modes being dependent on the mode order. Subsequently, the wave function overlaps of excitons and cavity photons changes, resulting in a variation of coupling strengths, which is reduced for lower order modes.

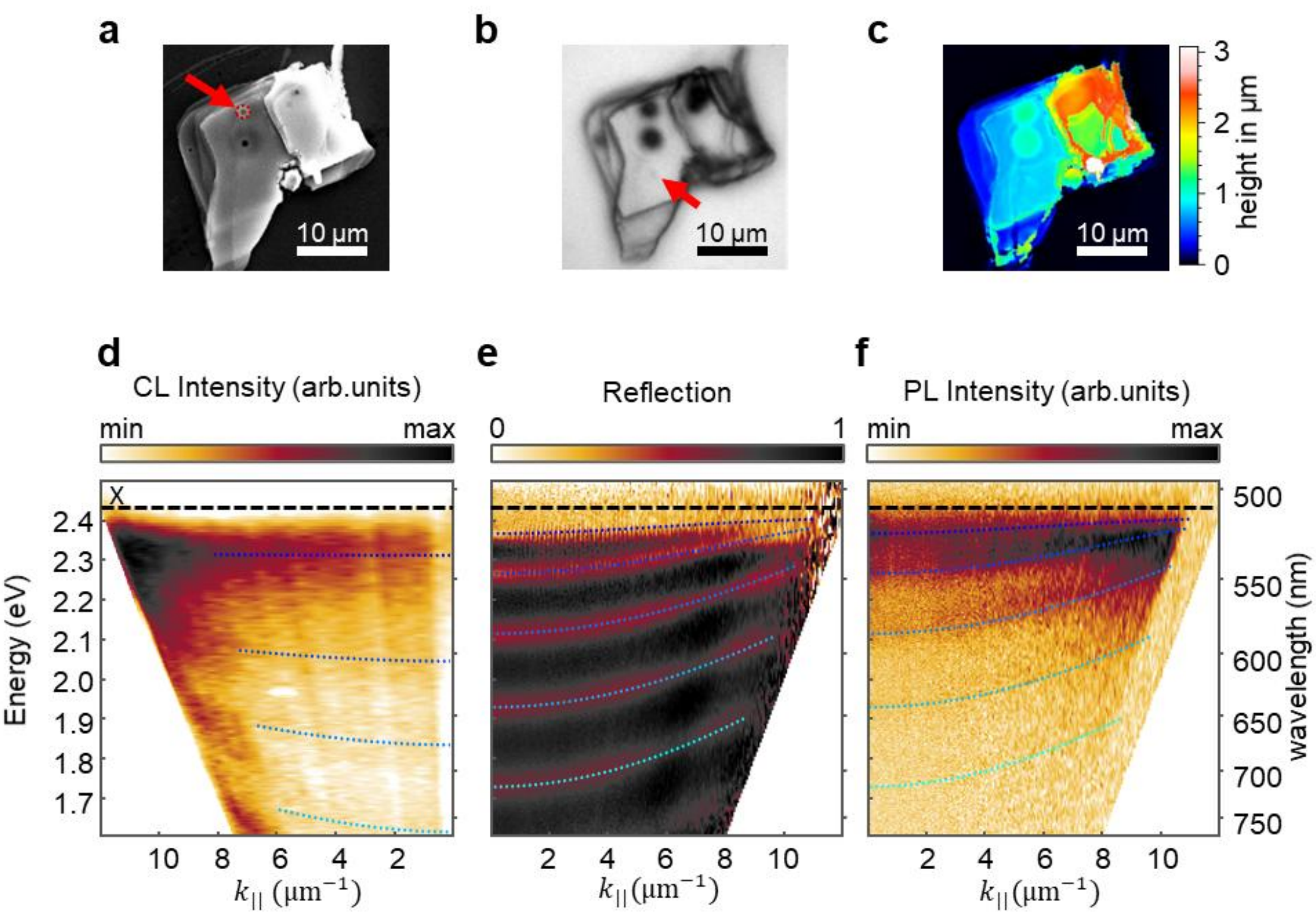

**Figure 3: Comparison of polaritonic dispersion relations under electron beam and optical excitation.** (**a**) SEM image and (**b**) optical reflection image of an RPP flake of 808 nm thickness with red arrows marking the position of the cathodoluminescence measurement in (a) and that of the optical measurements in (b). (**c**) Thickness map of the RPP flake determined by confocal optical profilometry. (**d** **e**, **f**) Momentum-resolved energy-momentum maps of cathodoluminescence at an initial kinetic energy of 20 keV for the electron beam, optical reflection and photoluminescence at $480\ \mathrm{nm}$ excitation wavelength, respectively. The Fabry–Pérot polaritonic dispersions of valleys in the reflectivity and peaks in cathodo- and photoluminescence are fitted using the Hopfield model and marked by dotted lines color-coded by Fabry–Pérot mode orders. While the modes positions are the same in the optical measurements in (e) and (f), the Rabi-splitting and polaritonic dispersion flattening is considerably larger under electron-beam irradiation due to an enhanced and simultaneous exciton excitation.

As mentioned above, in the momentum-resolved CL spectrum, two excitonic branches coexist, namely, Fabry–Pérot modes near the Γ point, as well as the guide modes near the light line. The Fabry–Pérot modes, whose fits are marked by dotted lines in Figure 3d, are flattened and have a red shift in their energy level. As discussed, it is expected that when multiple excitons are coherently excited and couple to one cavity mode, the level repulsion increases. Since a high energy electron is capable of exciting simultaneously multiple excitons in a confined volume much smaller than the wavelength of the light, the excitons are excited in-phase. Therefore, the substantially enhanced level repulsion and the corresponding mode shift and flattening originates from a larger number of excitons coupling simultaneously to a single cavity photon. By quantifying the level repulsion of the Fabry–Pérot part of the CL measurement, we report the coupling strength $g_{\mathrm{CL}} = 355\ \mathrm{meV}$ for the higher order LPBs, which is considerably larger than the retrieved coupling strength in the optical measurements. The ratio of the number of excitons coupling collectively per electron excitation compared to optical excitations can be therefore estimated by $N_{\mathrm{coh}} = \left(\frac{g_{\mathrm{CL}}}{g_{\mathrm{opt}}}\right)^2$ which shows that approximately 23 excitons are coherently excited under 20 keV electron beam irradiation.

The instability of halide perovskites even in relatively more stable 2D perovskites has to be taken into account in order to validate the observed phenomena[50]. Therefore, in Supplementary Fig. S6, the degradation effect has been investigated with continuous laser excitation. Over time, the PL spectra become less intense and feature a blue shift of the distinguishable modes, while the superposition of the higher-order LPBs remains at the same wavelength, an effect explainable by the flake thinning down. Contrary to this, the CL degradation in Fig. 3c shows no thinning, but rather a local elevation of the surface. Hence, the observation of fewer polaritonic modes in our detection range is not explainable by a thinning effect – but happens due to an increase in level repulsion due to a larger coupling strength.

For higher thicknesses of the RPP flakes, the differences between the electron-beam-induced and the laser-induced polarizations become even more drastically apparent. The CL, reflection and PL measurements for an RPP flake with the thickness of approximately 1300 nm demonstrate this fact (Fig. 4a to f). The thickness map was acquired using atomic force microscopy (AFM) (Fig. 4b), where by taking a line profile, the thickness was determined to be 1300 nm (Fig. 4c). Since a thicker flake hosts many Fabry–Pérot resonances in the relevant wavelength range, a large number of LPBs appear in the optical measurements, which converge with increasing mode orders to an intense PL peak at a fixed energy of 2.37 eV ($\lambda = 523\,\mathrm{nm}$). The superposition of the higher-order modes leads to an enhanced absorption efficiency and causes a cut-off of the reflectivity at a far lower energy than the exciton resonance. Using the Hopfield-fit method described in Supplementary section 2, we determine an average coupling strength of $g_{\mathrm{opt}} = 222\ \mathrm{meV}$, being substantially larger than the coupling strength for a flake with 808 nm thickness explored above. For a thicker flake, the number of quantum wells stacked to form such a flake increases substantially which effectively increases the number $N$ of available excitonic oscillators to couple to the photonic modes. Hence for thicker flakes, one observes as well collective excitations of excitonic oscillators to the same optical mode, compared to a thin flake even under optical excitation.

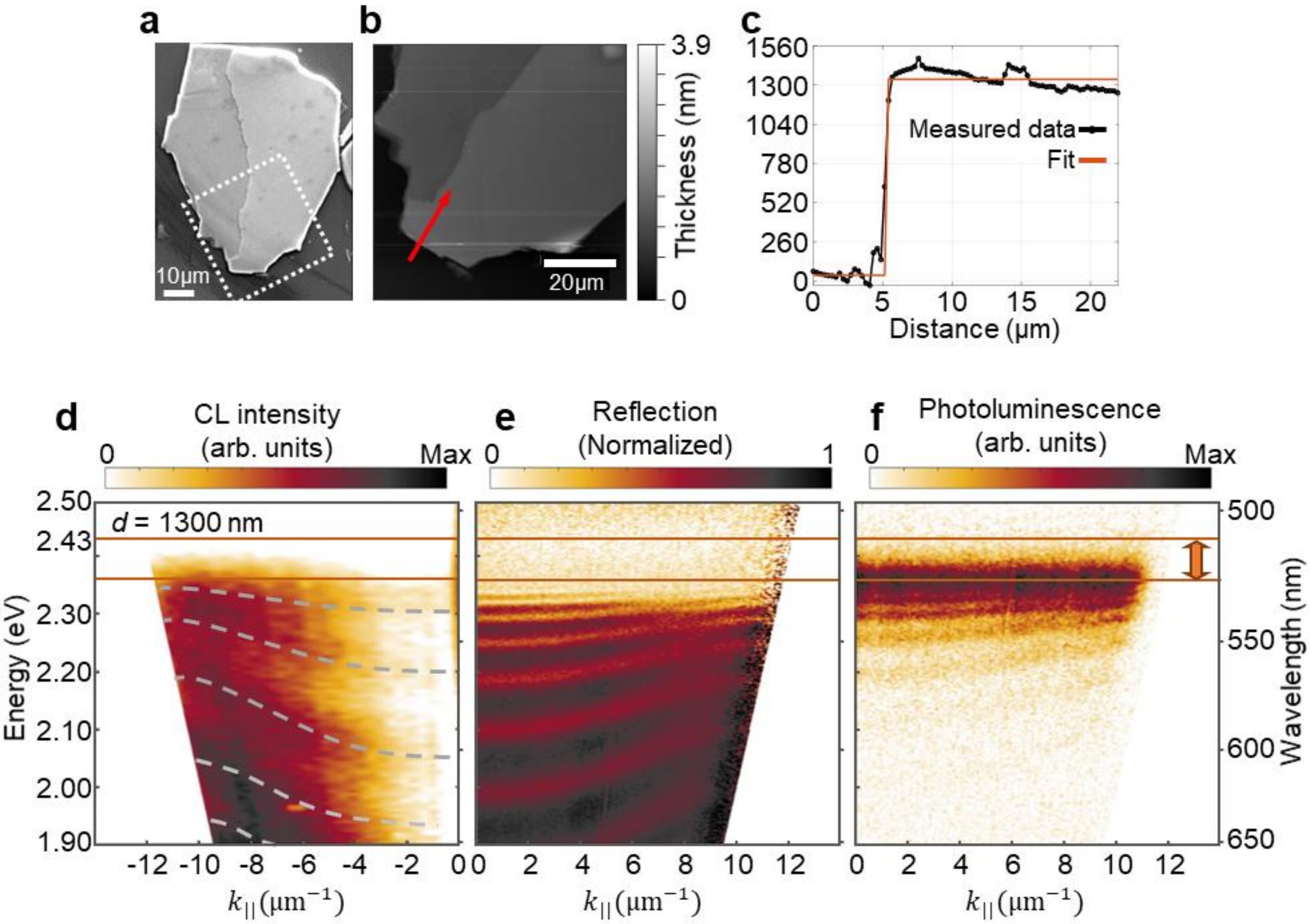

**Figure 4: Comparison between electron-based and optical excitation of a thick RPP layer.** (**a**) SEM image and (**b**) AFM image of a 1300 nm thick RPP flake, with the white dotted rectangle in (a) marking the area depicted in (b). **c**, Height profile along the red arrow in (b). (**d**) Momentum-resolved CL spectrum of the flake excited with an electron beam with the kinetic energy of 20 keV featuring polaritonic modes. (**e**, **f)** Momentum-resolved reflection and PL spectra with polaritonic modes appearing as dips in the former and peaks in the latter. A pronounced level repulsion is observed, apparent in (f) by a large shift of the superposition of multiple high-order modes.

A contrasting observation is found in our CL measurements of the same RPP flake in Fig 4d. While the red shift of the cut-off of the CL signal is similar to optical measurements, the fits of the peak dispersions (marked by grey-dotted lines) does not produce physically meaningful coupling strengths. Here, the coupling strengths strongly varies between 88 meV and 182 meV for each mode fitted separately, prompting the conclusion that the fitting model doesn't cover the optical response in the CL measurements properly. This is closely related to a transition to an incoherent excitation: When the electron beam traverses the thicker RPP, even an electron at the kinetic energy of 20 keV substantially loses its kinetic energy, excites secondary electrons with varying energies, thereby enlarging the interaction volume significantly. The latter decreases the spatial coherence of the excited excitons, while the strong deceleration leads to the diminishing of the phase relations between excited excitons. Thus, the electron-driven collective and coherent excitation of excitons is expected to be weakened with increasing the thickness, rather than being enhanced, in contrast to the optical excitations.

Arguably, this non-deterministic change of the excitation mechanism from thin to thick layers causes the CL response to become a complex composition of several coherent and incoherent excitation schemes, complicating its decomposition. We aim to quantify the transition from a synchronous to an asynchronous excitation scheme in the following section, but instead of varying RPP layer thickness, which would include changing the energy levels and field distribution of the cavity modes, we measure the CL response for varying initial electron kinetic energies, thereby controlling their relative energy loss and the spread of their trajectories.

**From synchronous to asynchronous excitation scheme with electron beams**

In the following, we demonstrate that a collective and coherent excitonic excitation underpins the formation of exciton polaritons in RPP under electron beam radiation. To demonstrate this, we analyze the transition to an asynchronous excitation scheme, which is realized by probing the CL response of an RPP flake to electron beams with kinetic energies varying from 20 keV to 12 keV.

Angle-resolved CL is performed on a 615 nm thick RPP layer positioned on 40 nm thick gold substrate with acceleration voltages of 12 kV, 16 kV, and 20 kV. While at a kinetic electron energy of 20 keV, a clear polaritonic dispersion relation featuring multiple LPBs is evident (Fig. 6a), the signal gets weaker at 16 keV and at 12 keV, and the transition radiation at 2.4 eV becomes dominant over the polaritonic radiation for the latter electron kinetic energy. Additionally, the polaritonic modes change both in spread and shape, indicating a weakening of the coupling strength, which is not possible to quantify due to weak signal-to-noise ratio and an overlap with the guided-wave modes at high collection angles. These observations showcase the transition from a synchronous excitation scheme of excitons to an asynchronous scheme, further underlined by the expectation that a decrease in collective coupling would minimize the intensity of polaritonic radiation.

To better explain this behaviour, Monte-Carlo simulations were conducted for a vacuum-RPP-gold-glass layered system resembling the sample conditions. The electron trajectories, including secondary and back-scattered electrons, excited by an electron beam with normal incidence from the vacuum side are calculated, showing low relative kinetic energy loss at 20 keV, leaving many of the electrons fast enough to pass the gold layer (Fig. 6d). The gradual change to 16 keV and to 12 keV in causes the electrons to lose larger relative parts of their kinetic energy inside the RPP layer, leading to more strongly deflected trajectories and substantial electron-beam broadening (Fig. 6e and f) and increasingly longer time intervals between excitations. We note that some high-kinetic-energy electrons are largely deflected in Figure 6d, apparently indicating a large spread. However, all Monte-Carlo simulations contained the same number of time steps, and thus, fast deflected electrons were able to travel further and experience more interactions, thereby agreeing with the argument of small temporal interval variations between excitations. The majority of electrons contributing to the collective coherent excitations at 20 keV remain in the smaller, conical interaction volume. These results showcase the main reasoning for enhanced collective coupling of excitons to the cavity in CL: Reducing the electron velocity leads to a larger interaction volume, longer interaction time inside that volume and thus an elongation of excitation intervals. Consequently, the excitons are excited less coherently, which results in decreased collective coherent coupling to the cavity mode and explains the observation of a polariton-dominated to a transition-radiation-dominated CL spectrum.

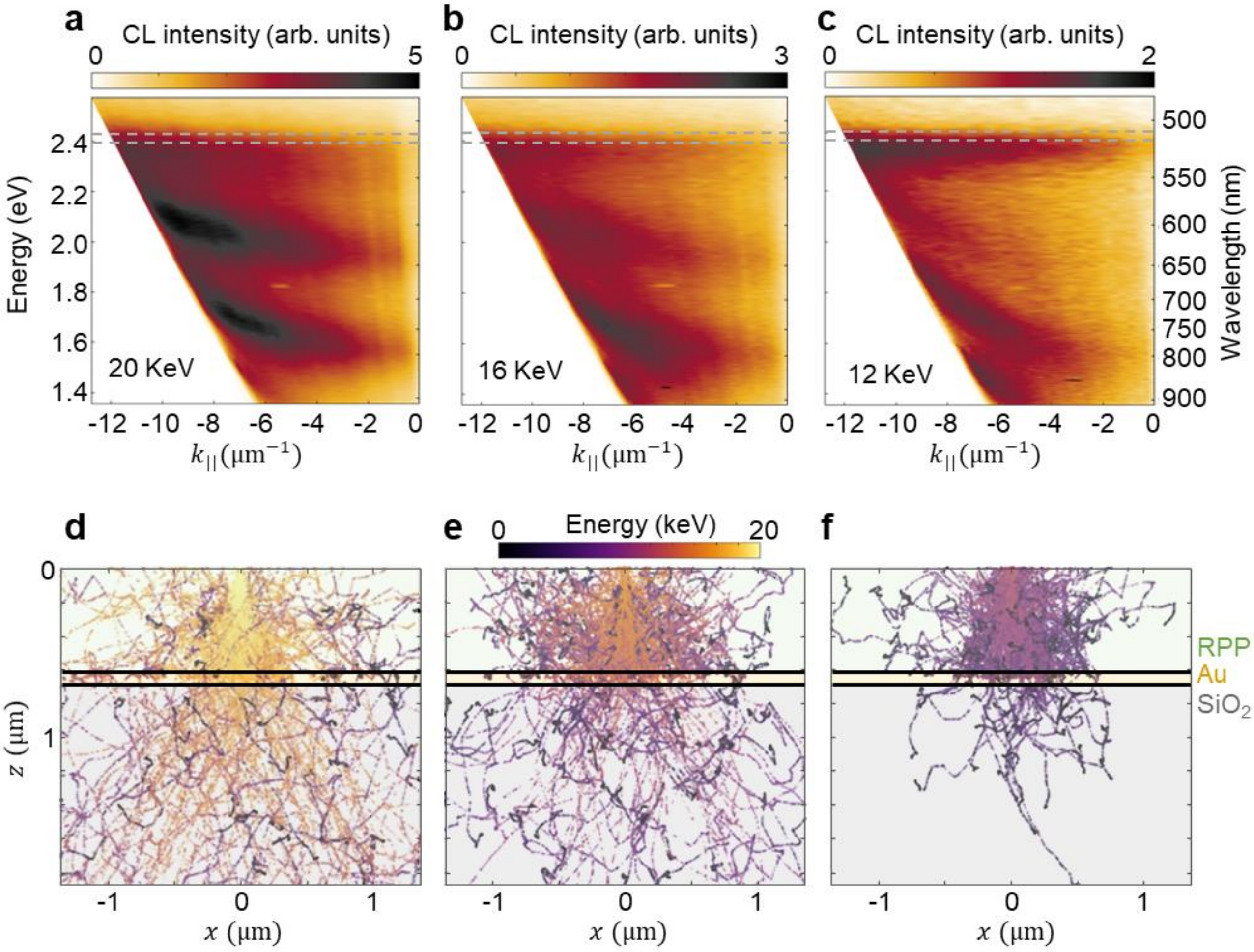

**Figure 5: Transition from synchronous to asynchronous electron beam excitation.** Momentum-resolved cathodoluminescence spectra of an RPP flake with 615 nm thickness acquired with **(a)** 20 kV, **(b)** 16 kV and **(c)** 12 kV acceleration voltages. The Fabry–Pérot polaritonic signal and the overall intensity decreases with lowering the electron kinetic energy, while the influence of the transition radiation becomes dominant. (**d**, **e**, **f)** Monte-Carlo simulations of electron trajectories and energies at acceleration voltages corresponding to (a), (b) and (c), respectively.

## Conclusion

In summary, we have investigated the fundamental differences of exciton-polariton formation between optical and electron beam excitation in Ruddlesden-Popper perovskite films on gold substrate. By correlating cathodoluminescence and photoluminescence maps, we exploited the strong-coupling effects between excitons and photons in RPP flakes as an exploratory laboratory for understanding how electron beams interact with quantum materials within the strong-coupling regime. By decreasing the electron velocity, we have shown a systematic change from polaritonic emission dominating the cathodoluminescence to the transition radiation being the main contribution, and we have linked this to a transition from coherent to incoherent excitonic excitations. Furthermore, evidence of single electrons exciting multiple excitons simultaneously was found, leading to enhanced polaritonic emission as well as an enhanced coupling strength compared to optical measurements on the same system, albeit for thinner RPP flakes. The increase of the coupling strength is explained by a higher number of excitonic oscillators coupling to the same photonic cavity mode, an effect that can be understood through multiple excitonic states being excited coherently, in-phase, and simultaneously, thereby acting as a coherent superposition that couples to the photonic mode. This phenomenon effectively multiplies the amount of independent identical excitonic oscillators by a factor of 23 in our measurements. Strikingly, we have correlated the transition from an asynchronous to a synchronous exciton excitation scheme, where the electron trajectories yield narrower interaction volumes and shorter intervals between excitations for fast electrons, thereby indicating a larger coherence of the excitations.

Our results demonstrate that synchronous excitation with fast electrons enhances and modifies the exciton polariton formation, solidifying our hypothesis of collective coupling of coherent excitons to photonic cavity modes. Next to the recent demonstration of super-absorption in confined volumes of perovskites due to the collective excitation of excitons[54], the observed giant enhancement of the coupling strength here, establishes perovskites as an exploratory platform for exploring light-matter interactions. Our study therefore provides important insight into the excitation mechanics of exciton polaritons under electron beam irradiation and opens up pathways of tuning of their dispersions. Therefore, this work will spark further investigations into high exciton-density mechanisms in similar systems and engineering the strength of light-matter interactions in active optoelectronic devices.

## Methodology:

### Synthesis of Ruddlesden-Popper Perovskites:

The $BA_2PbI_4$ was synthesized by reacting n-butylammonium iodide (BAI) and Lead(II) iodide ($PbI_2$) inside of a nitrogen-filled glovebox. The BAI was achieved by slowly adding 5 mL of 57% w/w aqueous hydriodic acid (HI) to 1 mL of n-$CH_3(CH_2)_3NH_3$ (BA) in an ice bath and stirring for 4 hours[55,56]. This solution was then transferred inside the glovebox. To achieve $PbI_2$, 500 mg PbO powder was dissolved in a mixture of HI (57%,3 ml) and $H_3PO_2$ (50%,850 µl) solution at the temperature of 130°C and stirring speed of 300 rpm for about 5 minutes to obtain a bright yellow solution of $PbI_2$. Thereafter, 1.5 ml of the synthesized BAI was added to the prepared PbI2 solution while increasing the solution temperature to 150°C and increasing the stirring speed to 400 rpm to achieve a clear solution. After 5 minutes in this condition, stirring and heating is stopped, letting the solution to cool down to room temperature, during which the orange bulk crystalline $BA_2PbI_4$ are emerged within the solution. Using vacuum filtering, $BA_2PbI_4$ products are separated and dried within the duration of a few days. The samples were fabricated by mechanically exfoliating RPPs on top of the substrates. The substrate for optical measurements on extremely thin flakes was glass (Fig. 2b). The substrate for cathodoluminescence measurements and relating optical measurements were a glass substrate with 40 nm gold on top of 10 nm titanium which was deposited using PVD. To achieve flakes with reduced thicknesses, a $BA_2PbI_4$ flake was picked up by a piece of scotch tape and after folding and unfolding for several times several flakes with reduced thicknesses were achieved. These flakes were stamped on Ar-plasma-treated substrates.

**Cathodoluminescence spectroscopy:**

The cathodoluminescence measurements were performed inside a ZEISS Sigma field-emission scanning electron microscope, equipped with the Delmic SPARC CL system. Throughout these measurements the acceleration voltage of the electron-beam varies from 10 kV to 20 kV, whereas the beam current has been kept at minimum current that the system allows (around 200 pA) to hinder electron-beam-induced degradation in RPPs, unless otherwise specified. The CL detector consists of an off-axis paraboloid mirror which is aligned with an accuracy less than $1 \mu m$ above the specimen and redirects the colected CL radiation towards the analyzing path including spectrographs, required optical filtering and alignment elements, and a CCD camera (Andor i-KonM). The parabolic mirror has an acceptance angle of $1.49 \pi sr$ $(NA = 0.97)$ and a focal distance of $0.5$ mm. The exciting electron-beam can pass the mirror through a hole with a diameter of $600 \mu m$ located above the focal point. Utilizing a dispersive reflection grating and on occasion directing the CL radiation through a one-dimensional slit opening, the CL detector is capable of performing energy-momentum imaging. The acquisition time for each pixel was set to values between 30 s and 120 s for energy-momentum mapping.

**Reflection and photoluminescence spectroscopy:**

The optical spectroscopy measurements were conducted using a Nikon Eclipse Ti2-A inverted microscope combined with a Princeton Instruments HRS-500 grating spectrometer in a system provided by the New Technologies and Consulting Company. Between the microscope and the spectrometer, a 4f relay lens system with an insertable Bertrand lens—effectively halving the focal length of the second relay lens—enabled switching between the back focal plane and the image being projected onto the spectrometer slit. Inside, the spectrometer slit image was split into its spectral components by a reflective grating with 150 lines per mm. All optical measurements we're performed with an objective with a magnification of 100x and a numerical aperture $NA = 0.9$, which limited the maximum collected angle to 64.158° with respect to the surface normal. The reflection spectra of the RRP layers on gold were recorded with a broadband halogen lamp as light source and were normalized to spectra of pure gold under similar conditions. As illumination source for the for the photoluminescence, a supercontinuum white-light laser (SC-OEM by YSL Photonics) was filtered by an acousto-optic tunable bandpass filter (AOTF-PRO-D by YSL Photonics) and focused onto the sample through the microscope objective itself, determining the smallest achievable laser spot size to be diffraction limited to 325 nm. Therefore, the illumination and the emission of the sample both pass the objective, and consequently, they had to be separated by a epifluorescence filter-cube featuring a 492 nm short-pass as excitation filter and a 500 nm long-pass as emission filter. To gain appropriate signal-to-noise ratios in all optical measurements, the acquisition times were varied between 2 s and 30 *s.*

**Monte Carlo simulation:**

Monte Carlo simulations were performed using the CASINO v2.42 package to investigate electron trajectories and energy dissipation within the multilayer structure[51]. The layer thicknesses were defined according to the fabricated sample geometry, with an RPP density of 2.676 g $cm^{-3}$. A vertically incident electron beam with a diameter of 10 nm and 10000 incident electrons was simulated at acceleration voltages of 12, 16, and 20 kV. The software simulates elastic scattering events and approximates inelastic interactions based on the mean energy loss between two successive elastic-scattering events[52,53]. The Mott approximation[54] was used to interpolate the total elastic-scattering cross-section, while the Joy–Luo method[52] was applied to account for the ionization potential, electron collisions, and inelastic energy losses.

**Acknowledgement**

This project received funding from the Volkswagen Foundation (Momentum Grant), European Research Council (ERC Consolidator Grant UltraSpecT with no. 101170341; ERC Proof-of-Concept Grant UltraCoherentCL with no. 101157312) and from Deutsche Forschungsgemeinschaft (Grant no. 554905035 and 447330010).

# Supporting Information for:
# Electron-beam-driven collective self-hybridized exciton polaritons

Parsa Darman[1ǂ]*, Maximilian Black[1ǂ]*, Prabhdeep Singh[1], Victor DeManuel-Gonzales[1], Sara Darbari[2], Masoud Taleb[1,3], Fatemeh Chahshouri[1,3], Nahid Talebi[1,3]*

1 Institute of Experimental and Applied Physics, Kiel University, 24418 Kiel, Germany

2 Faculty of Electrical and Computer Engineering, Tarbiat Modares University, Tehran 1411713116, Iran

3 Kiel Nano, Surface, and Interface Science KiNSIS, Kiel University, 24118 Kiel, Germany

[ǂ] These authors contributed equally to this work.

Email: talebi@physik.uni-kiel.de, darman@physik.uni-kiel.de, black@physik.uni-kiel.de

**Content:**

**Supplementary Note S1: Hopfield model, Tavis Cummings model, and the determination of the coupling strength**

To extract the coupling strength $g$ for the measured polaritonic dispersions in the cathodoluminescence and optical measurements, we fit the polaritonic modes, i. e. the lower polaritonic branches (LPB), with energies derived from the Hopfield model[1] and its extension for $N$ excitonic oscillators, the Tavis-Cummings model[2].

In the rotating wave approximation and with neglection of diamagnetic terms, the Hopfield Hamiltonian for an exciton interacting with a photonic mode is:

$$\hat{H} = \hbar\omega_c \hat{a}^\dagger \hat{a} + \hbar\omega_e \hat{b}^\dagger \hat{b} + \hbar g\left(\hat{a}^\dagger \hat{b} + \hat{b}^\dagger \hat{a}\right) \qquad \text{(S1)}$$

Here, $\hat{a}^\dagger$, $\hat{a}$ and $\hat{b}^\dagger, \hat{b}$ are the creation and annihilation operators for the photon and the exciton, respectively, $\omega_c$ is the cavity photon frequency and $\omega_e$ denotes the exciton frequency. From this, the Eigenvalues, and therefore the polaritonic energies $E_{\mathcal{H}}$, are derived:

$$E_H = \frac{1}{2}\hbar\left(\omega_c + \omega_e\right) \pm \frac{1}{2}\hbar\sqrt{\left(\omega_c - \omega_e\right)^2 + 4g^2} \qquad \text{(S2)}$$

$\omega_e$ is given, $E_{\mathcal{H}}(k_{||})$ is measured and $g$ needs to be a fit parameter, therefore $\omega_c(k_{||})$ of the Fabry–Pérot modes is to be determined for each LPB separately with mode order $m$. The cavity is formed by the parallel surfaces of a crystal hosting excitons and $E_{\mathcal{H}}(k_{||})$ is measured within the light cone, where the Fabry–Pérot modes interact with the excitons. Fabry–Pérot modes yield the following condition ($L$: cavity length, $\phi$: phase shift at internal reflection, $m$: mode number):

$$2k_\perp L + 2\phi = 2m\pi \Leftrightarrow k_\perp = \frac{m\pi}{L + \delta L} \qquad \text{(S3)}$$

The phase shift $2\phi = \delta L \cdot k_\perp$, which originates from the internal reflection at the surfaces of the RPP crystal, can be viewed as a shift in the optical path length, since this would cause the same change. Hence, we define and calculate $\omega_c$:

$$\omega_c = ck_{0,c} = c\sqrt{\frac{k_{||}^2 + k_\perp^2}{\varepsilon_{RPP,b}}}, \qquad \text{(S4)}$$

where $k_{||}$ and $k_\perp$ are the in-plane and out-of-plane components of the wave vector of the cavity photon and $c$ is the velocity of light. With this, the energies to be fitted become $E_{\mathcal{H}} = E_{\mathcal{H}}(k_{||})$ for a given mode order $m$, and the fitting parameters are $g$ and $\delta L$. The $BA_2PbI_4$ permittivity[1] is frequency-dependent, and to solve $\omega_c(k_{||})$, an expression for the background dielectric function $\varepsilon_{\mathrm{RPP,b}} = \varepsilon_{\mathrm{RPP,b}}(\omega_c)$ is needed. Since the exciton is taken into account in the Hopfield Hamiltonian, $\varepsilon_{\mathrm{RPP,b}}$ has to reflect the dielectric background that the exciton experiences. To solve this, we fit the real and the imaginary part of the measured RPP permittivity[1] with the sum of two Lorentzian functions and a quadratic background function, which is properly reproducing the dielectric function in Supplementary Fig. S1a. The first Lorentzian is the footprint of the exciton itself. The exciton-independent dielectric function becomes

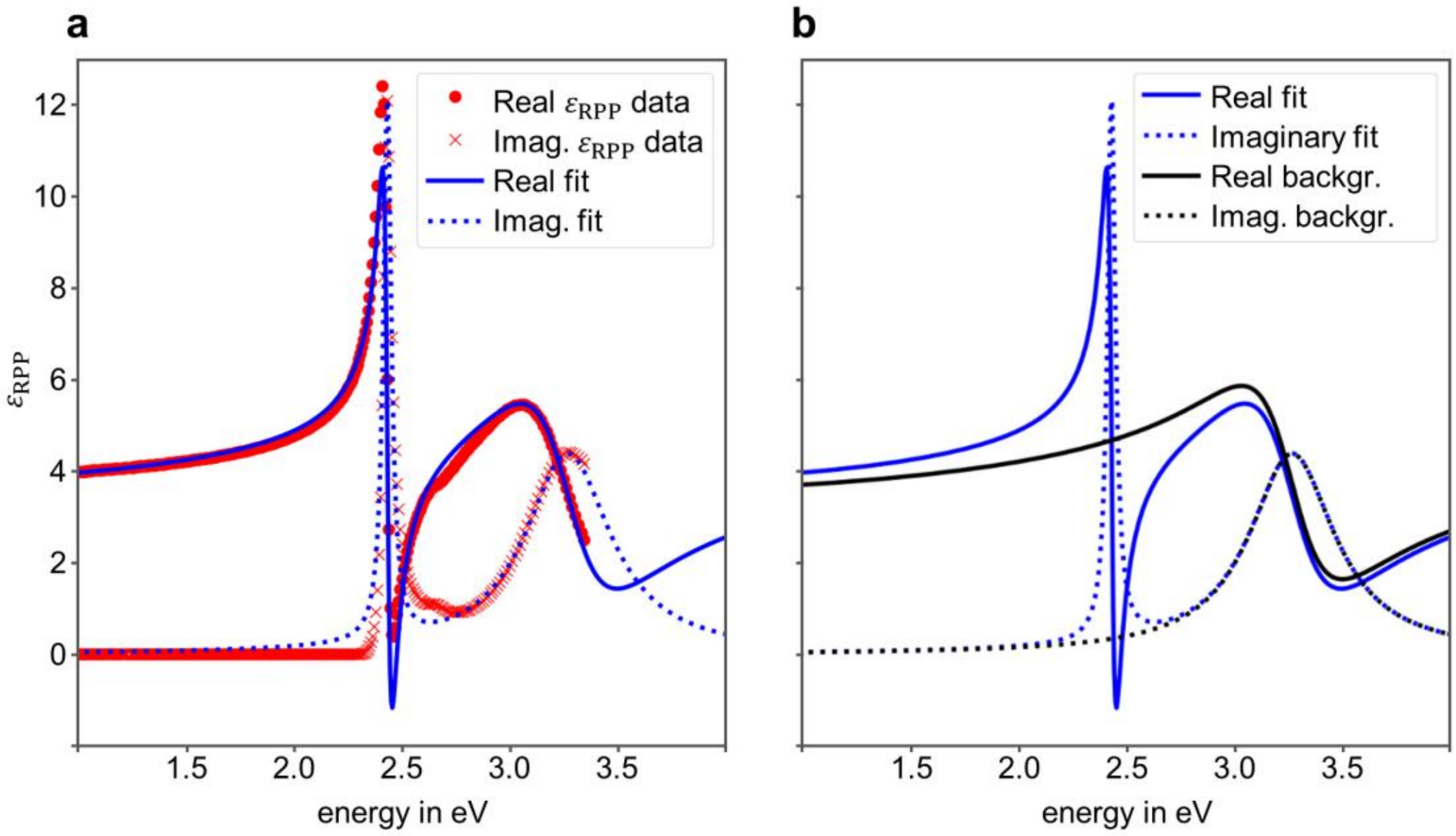

**Supplementary Figure S1. Dielectric background of $BA_2PbI_4$. (a**) Fitted real and imaginary parts of the dielectric function depicted as solid and dashed blue lines, respectively, showcasing that the permittivity can be reconstructed as the sum of two Lorentzian functions and a nearly constant function. The experimentally retrieved real and imaginary parts of the permittivity are depicted with red dots and crosses, respectively. (**b)**, The dielectric background without the exciton (black), resulting when the excitonic Lorentzian is subtracted from the fitted permittivity (blue), thereby consisting of a Lorentzian above the exciton energy plus a nearly constant function.

$$\varepsilon_{\mathrm{RPP,b}}\left(\omega_c\right) = a\omega_c^2 + \varepsilon_{\mathrm{inf}} + \frac{A}{\omega_0^2 + \omega_c^2 - i\gamma\omega_c}, \tag{S5}$$

with fitting parameters $a$, $\varepsilon_{\mathrm{inf}}$, $A$, $\omega_0$ and $\gamma$. Thus, equation S3 becomes:

$$a\omega_c^4 + \varepsilon_{\mathrm{inf}}\,\omega_c^2 + \frac{A\omega_c^2}{\omega_0^2 - \omega_c^2 - i\gamma\omega_c} - c^2k^2 = 0\,, \tag{S6}$$

with $k_{||}^2 + k_{\perp}^2 = k^2$. This is solved numerically for $\omega_c$ with each $k_{||}$ value. We note that equation S6 has multiple complex solutions for $\omega_c$, but only those with positive real and small negative imaginary part make sense physically and are thereby taken into account. Due to the complex background $\varepsilon_{\mathrm{RPP,b}}$, $\omega_c$ becomes complex as well. While the Hopfield Hamiltonian in general is loss-free, for weak damping and for coupling below the ultrastrong coupling regime, the damping of the photon and exciton can phenomenologically be take into account in equation S1 by replacing $\omega_c$ with $\widetilde{\omega}_c = \omega_c - \frac{\mathrm{i}\gamma_c}{2}$, and similarly, $\omega_e$ with $\widetilde{\omega}_e = \omega_e - \frac{\mathrm{i}\gamma_e}{2}$. [2-4] Here, $\gamma_c$ is found by solving equation S6 and $\gamma_e$ is determined by the Lorentzian fit of the $\varepsilon_{\mathrm{RPP}}(\omega)$ data. Now, with $\widetilde{\omega}_c(k_{||})$ known for each $k_{||}$, the LPBs are extracted from the energy-momentum maps as peaks in the cathodoluminescence and photoluminescence and as dips in the reflection data, and $E_{\mathcal{H}}(k_{||})$ is fitted to them. For the fit, it is important to take into account that the coupling strength $g$ and the effective optical path length change $\delta L$ are anticorrelated, since both shift the energy level of a mode and change its curvature. This makes a proper choice of starting parameters and bounds especially mandatory, especially for $\delta L$. Following, our choices are reasoned and the initial parameter plots are shown in Supplementary Fig. S2.

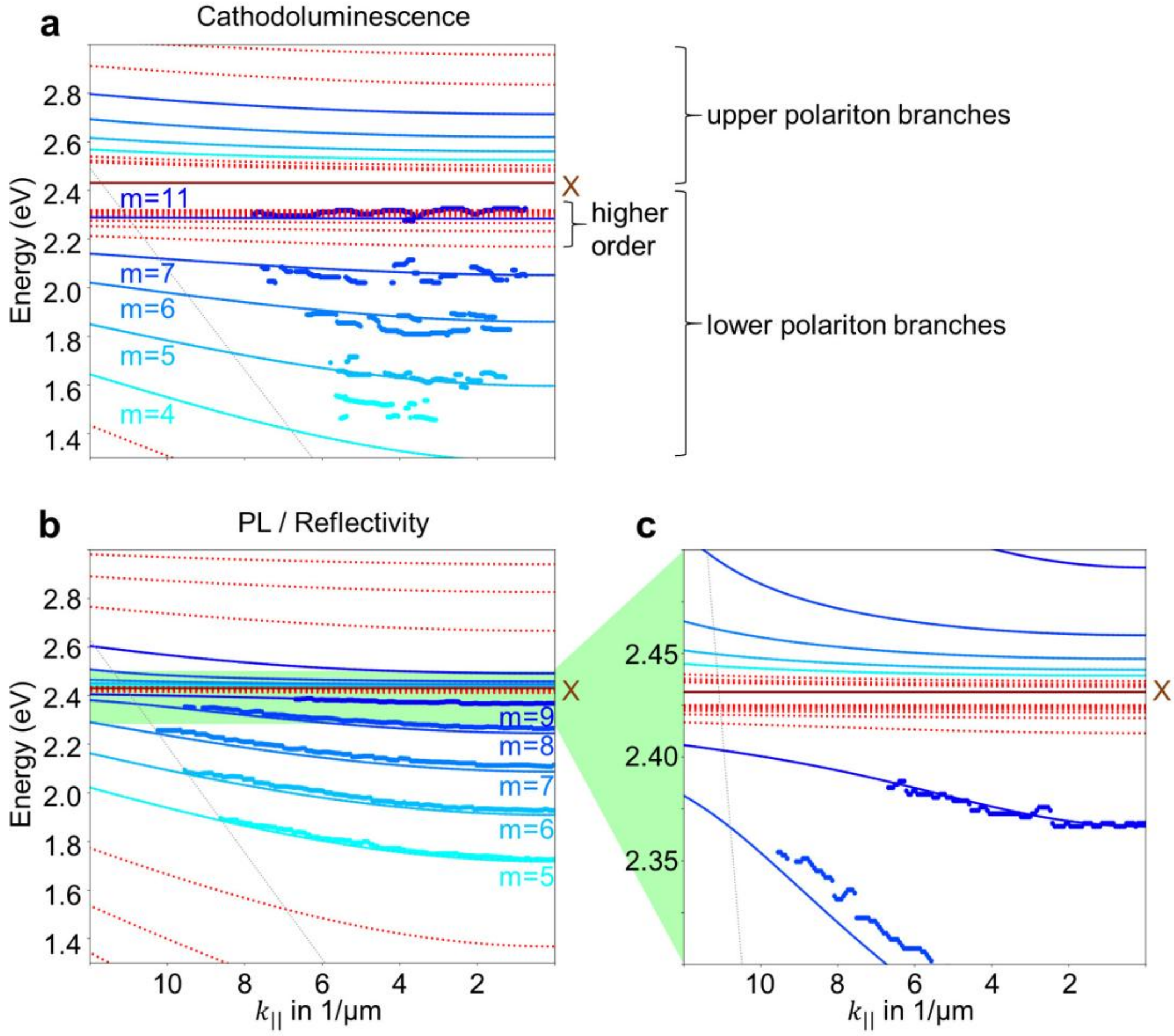

**Supplementary Figure S2. Determination of the coupling strength under electron and optical excitation. (a)** Dispersion relations of lower polaritonic modes plotted with the initial parameters of the fit to the peaks in the CL data in Figure 3d in the main text. The cyan to blue colour coded data points denote peaks attributed to a mode order *m*, while the similarly coloured lines are the fitted lower polaritonic modes and—above the exciton energy—their corresponding upper polariton branches. Red dotted lines showing polaritonic modes calculated for multiple mode orders demonstrate the convergence of higher-order lower polariton modes to a constant energy level. (**b**) Initial parameters plot of the lower polaritonic branches in the response of the RPP to optical excitation. (**c**) Enlarged energy-momentum map corresponding to the green-marked area in (b), demonstrating calculated higher-order LPBs (red-dotted lines) below the excitons approaching a constant energy level, whose closer proximity to the exciton resonance is caused by a smaller coupling strength compared to CL.

In the CL data of Fig. 3 in the main text, it is evident that the signal close to the exciton resonance is comprised of multiple modes. To get a reliable value for the coupling efficiency, we use an ansatz for high-order polariton branches. For high-order polaritonic modes, a good assumption is that the coupling strength becomes constant, since the change in electric field distribution between two neighbouring high-order Fabry–Pérot modes, and thereby the overlap between photonic and excitonic mode, is comparably small. This is fairly understandable due the spatial periods of the standing wave inside the RPP layer, which are already large, being increased by only 1. Given a constant coupling strength for high-order modes, the energy of the higher-order LPBs approach a constant value, which is red-shifted with respect to the exciton energy. This is visualized by plotting the calculated higher-order modes for CL in Supplementary Fig. S2a and for optical measurements in Supplementary Fig. 2c. This convergence results from equation S7:

$$\lim_{m\to\infty} k_\perp = \infty \Rightarrow \lim_{m\to\infty} c^2k^2 = \infty \Rightarrow \lim_{m\to\infty}\left(a\omega_c^4 + \varepsilon_{\inf}\omega_c^2 + \frac{A\omega_c^2}{\omega_0^2 - \omega_c^2 - i\gamma\omega_c}\right) = \infty \tag{S7}$$

Apart from the trivial solution $\omega_{c,m\to\infty} = \infty$, we get $\omega_{c,m\to\infty} = \sqrt{\omega_0^2 - \frac{\gamma^2}{4}} - \frac{\gamma}{2}\mathrm{i}$, and therefore a constant value for $E_{\mathcal{H}}$, from which $g$ is determined. This is a direct causal consequence of the Lorentzian in $\varepsilon_{\mathrm{RPP,b}}$. It is evident that $E_{\mathcal{H}}$ is insensitive to the phase shift for higher order modes, and consequently, the fitting parameters $g$ and $\delta L$ are not anticorrelated in this regime. Hence, for an arbitrary higher order mode, we determine the average coupling strength for the higher-order modes from the CL data. By choosing adequate mode numbers for the lower-order modes and limiting $\delta L$ to prohibit mode jumps by changing the optical path length, we additionally obtain estimates of the coupling strengths of the other observed modes in the CL data.

When multiple excitonic oscillators are excited, they can collectively couple to the optical cavity mode. This is described by the Tavis-Cummings model[5] with the vector basis $\{|C\rangle, |X_1\rangle, |X_2\rangle, \cdots, |X_N\rangle\}$, where $|C\rangle$ denotes the excitation of the cavity mode and $|X_i\rangle$ that of the *i*-th exciton. The Hopfield Hamiltonian is extended to

$$\hat{H}_{\mathrm{TC}} = \hbar\omega_c\hat{a}^\dagger\hat{a} + \hbar\omega_e\sum_{i=1}^{N}\hat{b}_i^\dagger\hat{b}_i + \hbar g\sum_{i=1}^{N}\left(\hat{a}_i^\dagger\hat{b}_i + \hat{b}_i^\dagger\hat{a}_i\right). \tag{S8}$$

A unitary transformation to a bright state $|\mathrm{B}\rangle = \frac{1}{\sqrt{N}}\sum_{i=1}^{N}|\mathrm{X}_i\rangle$ and *N* orthogonal dark states $|\mathrm{D}_k\rangle$ yields

$$\hat{H}_{\mathrm{TC}} = \hbar\omega_c\hat{a}^\dagger\hat{a} + \hbar\omega_e\hat{B}^\dagger\hat{B} + \hbar g\sqrt{N}\left(\hat{a}^\dagger\hat{B} + \hat{B}^\dagger\hat{a}\right) + \hbar\omega_e\sum_{k=1}^{N-1}\hat{D}_k^\dagger\hat{D}_k\,. \tag{S9}$$

Here, $B^\dagger$ and $B$ denote the creation and annihilation operators of the bright state and $D_k^\dagger$ and $D_k$ those of the dark states. By definition, the dark states do not couple to the cavity, making the last term in the Hamiltonian irrelevant for the polaritonic energies, and thereby $\mathcal{H}_{\mathrm{TC}}$ resembles $\mathcal{H}$ with $g_N = \sqrt{N}g$. Thus, an *N*-fold increase in excitonic oscillators will multiply the coupling strength by $\sqrt{N}$, allowing us to determine the factor *N*, by which more excitonic oscillators couple coherently to the photonic mode in CL as compared to optical excitation.

It is important to understand that the bright mode is inherently a coherent superposition of the excitonic states. More generally, one can write the bright state as $|B_{\mathrm{coh}}\rangle = \frac{1}{\sqrt{N}}\sum_{j=1}^{N}e^{i\theta_j}|X_j\rangle$ and the corresponding creation operator as $\hat{B}_{\mathrm{coh}}^\dagger = \sum_{j=1}^{N}e^{i\theta_j}\hat{b}_j$, where $\theta_j$ denotes a constant phase relation. In the case of nearly simultaneous excitation of the excitonic oscillators via the ultrashort interaction time with the electron beam, one can neglect the phase delay ($\theta_j \simeq 0$). Thus, in this case, the relation $g_{\mathrm{coh}} = \sqrt{N_{\mathrm{coh}}}g$ for $N_{\mathrm{coh}}$ coherently excited excitons is equivalent to $g_N = \sqrt{N}g$ for a system with *N* excitonic oscillators available for excitation.

**Supplementary Note S2: Analytic calculation of the cathodoluminescence radiation**

We consider an electron with the velocity $v_e$ moving along the z axis in the direction specified in Supplementary Fig. S3a. The current distribution function associated with this moving electron is specified as $\vec{J}(x,y,z;t) = -ev_e\,\hat{z}\,\delta(x)\delta(y)\delta(z+v_e t) = -ev_e\,\hat{z}\,\delta(z+v_e t)\iint dk_x\,dk_y\,e^{ik_x x}e^{ik_y y}$ , and its Fourier transform is revealed as

$$\tilde{\vec{J}}(x,y,z;\omega) = -\frac{ev_e}{2\pi}\hat{z}\int\delta(z+v_e t)e^{i\omega t}dt\iint dk_x\,dk_y\,e^{ik_x x}e^{ik_y y} = -\frac{e}{2\pi}\hat{z}\,e^{-i\frac{\omega}{v_e}z}\iint dk_x\,dk_y\,e^{ik_x x}e^{ik_y y}\ . \quad \text{(S10)}$$

Equation S10 should be inserted in the Helmholtz equation $\nabla^2\vec{A} + \varepsilon_r k_0^2\vec{A} = -\vec{J}$ (where we consider the convention $\vec{H} = \vec{\nabla}\times\vec{A}$), to calculate for the vector potential and respectively derive the field components. By considering a solution ansatz in the form $\vec{A}(x,y,z;\omega) = \iint dk_x\,dk_y\,\tilde{\tilde{\vec{A}}}(k_x,k_y,z)e^{ik_x x}e^{ik_y y}$ , the Helmholtz equation is simplified into a one-dimensional partial differential equation of the form

$$\frac{\partial^2\tilde{\tilde{\vec{A}}}(k_x,k_y,z)}{\partial z^2} + \left(\varepsilon_r k_0^2 - k_x^2 - k_y^2\right)\tilde{\tilde{\vec{A}}}(k_x,k_y,z) = e\,\hat{z}\,e^{-i\frac{\omega}{v_e}z}\ . \quad \text{(S12)}$$

The solutions for an electron propagating in a homogenous material system with the permittivity $\varepsilon_r$ is hence given by

$$\tilde{A}_z^{\mathrm{inc}} = \frac{e}{2\pi}\frac{1}{\varepsilon_r k_0^2 - k_x^2 - k_y^2 - (\omega/v_e)^2}e^{-i\frac{\omega}{v_e}z} = \frac{e}{2\pi\,\kappa_{\mathrm{inc}}}e^{-i\frac{\omega}{v_e}z}\ , \quad \text{(S13)}$$

where $\kappa_{\mathrm{inc}} = \varepsilon_r k_0^2 - k_x^2 - k_y^2 - (\omega/v_e)^2$. For the multilayer system with the geometry depicted in Supplementary Fig. S3a, the solution Ansatz for the vector potential is constructed as

$$\tilde{A}_z = \begin{cases} A_r^z e^{+ik_z^{(1)}z} + \dfrac{e}{2\pi\,\kappa_{\mathrm{inc}}^{(1)}}e^{-i\frac{\omega}{v_e}z} & z>0 \\[2ex] A_1^z e^{-ik_z^{(2)}z} + A_2^z e^{+ik_z^{(2)}(z+d)} + \dfrac{e}{2\pi\,\kappa_{\mathrm{inc}}^{(2)}}e^{-i\frac{\omega}{v_e}z} & -d<z<0 \\[2ex] A_t^z e^{-ik_z^{(3)}(z+d)} + \dfrac{e}{2\pi\,\kappa_{\mathrm{inc}}^{(3)}}e^{-i\frac{\omega}{v_e}z} & z<-d \end{cases} \quad \text{(S14)}$$

with $k_z^{(1)} = \sqrt{k_0^2 - k_x^2 - k_y^2}$ , $k_z^{(2)} = \sqrt{\varepsilon_{rd}k_0^2 - k_x^2 - k_y^2}$ , $k_z^{(3)} = \sqrt{\varepsilon_{rm}k_0^2 - k_x^2 - k_y^2}$ , $\kappa_{\mathrm{inc}}^{(1)} = k_0^2 - k_x^2 - k_y^2 - (\omega/v_e)^2$, $\kappa_{\mathrm{inc}}^{(2)} = \varepsilon_{rd}k_0^2 - k_x^2 - k_y^2 - (\omega/v_e)^2$, and finally

$\kappa_{\text{inc}}^{(3)} = \varepsilon_{rm}k_0^2 - k_x^2 - k_y^2 - (\omega/v_e)^2$. Here, $\varepsilon_{rd}$ is the permittivity of the 2D Ruddlesden-Popper-type halide perovskite material system (see Supplementary Fig. S3b), and $\varepsilon_{rm}$ is that of the gold substrate. The unknown amplitudes $A_r$, $A_2$, $A_3$, and $A_t$ should be solved using the tangential boundary conditions. The electric field and magnetic field components are given by

$$\vec{E} = i\omega\mu_0\vec{A} - \frac{1}{i\omega\varepsilon_0\varepsilon_r}\vec{\nabla}\vec{\nabla}\cdot\vec{A} = i\omega\mu_0 A_z\hat{z} - \frac{1}{i\omega\varepsilon_0\varepsilon_r}\vec{\nabla}\left(\frac{\partial A_z}{\partial z}\right), \tag{S15a}$$

and

$$\vec{H} = \vec{\nabla}\times\vec{A} = \vec{\nabla}A_z \times \hat{z} = +\frac{\partial A_z}{\partial y}\hat{x} - \frac{\partial A_z}{\partial x}\hat{y}, \tag{S15b}$$

Respectively. Obviously, the field profiles would be $TM_z$ polarized, expressing no *z*-component for the magnetic field. The field amplitudes $A_1$ and $A_2$ are obtained as

$$A_1 = \frac{1}{\Delta(k_{\|},\omega)}\left(\left[-\frac{k_z^{(3)}}{\varepsilon_{rm}} - \frac{k_z^{(2)}}{\varepsilon_{rd}}\right]E_1 - \left[k_z^{(1)} - \frac{k_z^{(2)}}{\varepsilon_{rd}}\right]e^{+ik_z^{(2)}d}E_2\right), \tag{S16a}$$

and

$$A_2 = \frac{1}{\Delta(k_{\|},\omega)}\left(-\left[-\frac{k_z^{(3)}}{\varepsilon_{rm}} + \frac{k_z^{(2)}}{\varepsilon_{rd}}\right]e^{+ik_z^{(2)}d}E_1 + \left[k_z^{(1)} + \frac{k_z^{(2)}}{\varepsilon_{rd}}\right]E_2\right), \tag{S16b}$$

with

$$\Delta(k_{\|},\omega) = \left[k_z^{(1)} + \frac{k_z^{(2)}}{\varepsilon_{rd}}\right]\left[-\frac{k_z^{(3)}}{\varepsilon_{rm}} - \frac{k_z^{(2)}}{\varepsilon_{rd}}\right] - \left[k_z^{(1)} - \frac{k_z^{(2)}}{\varepsilon_{rd}}\right]\left[-\frac{k_z^{(3)}}{\varepsilon_{rm}} + \frac{k_z^{(2)}}{\varepsilon_{rd}}\right]e^{+i2k_z^{(2)}d}, \tag{S17a}$$

$$E_1 = \frac{e}{2\pi}\frac{1}{\kappa_{\text{inc}}^{(1)}}\left(\frac{\omega}{v_e} + k_z^{(1)}\right) + \frac{e}{2\pi}\frac{1}{\kappa_{\text{inc}}^{(2)}}\left(\frac{1}{\varepsilon_{rd}}\frac{\omega}{v_e} - k_z^{(1)}\right), \tag{S17b}$$

and

$$E_2 = \frac{e}{2\pi}e^{+i\frac{\omega}{v_e}d}\left[\frac{1}{\kappa_{\text{inc}}^{(2)}}\left(-\frac{1}{\varepsilon_{rd}}\frac{\omega}{v_e} + \frac{k_z^{(3)}}{\varepsilon_{rm}}\right) + \frac{1}{\kappa_{\text{inc}}^{(3)}}\left(\frac{1}{\varepsilon_{rm}}\frac{\omega}{v_e} - \frac{k_z^{(3)}}{\varepsilon_{rm}}\right)\right]. \tag{S17c}$$

Consequently, $A_r$ and $A_t$ are further obtained as

$$A_r = A_1 + A_2 e^{+ik_z^{(2)}d} + \frac{e}{2\pi}\left(\frac{1}{\kappa_{\text{inc}}^{(2)}} - \frac{1}{\kappa_{\text{inc}}^{(1)}}\right),$$

and

$$A_t = A_1 e^{+ik_z^{(2)}d} + A_2 + \frac{e}{2\pi} e^{+i\frac{\omega}{v_e}d} \left( \frac{1}{\kappa_{\text{inc}}^{(2)}} - \frac{1}{\kappa_{\text{inc}}^{(3)}} \right).$$

For modelling the momentum-resolved CL spectra, we need to calculate the optical power transmitted to the far-field. We only consider the optical waves propagating towards +*z* direction, since our parabolic mirror is positioned above the sample, collecting the light propagating in the upward direction. We model this as the z-component of the Poynting vector integrated over an xy-plane at the far-field, specified as $P_z\left(k_x,k_y,\omega\right) = \frac{1}{2}\text{Re}\left\{\tilde{E}_x\left(k_x,k_y,\omega\right)\tilde{H}_y^*\left(k_x,k_y,\omega\right) - \tilde{E}_y\left(k_x,k_y,\omega\right)\tilde{H}_x^*\left(k_x,k_y,\omega\right)\right\}$, which is further obtained as

$$P_z\left(k_x,k_y,\omega\right) = \left|\tilde{A}_r^z\right|^2 \frac{\left(k_x^2+k_y^2\right)k_z^{(1)}}{\omega\varepsilon_0} = \left| A_1^z + A_2^z e^{+ik_z^{(2)}d} + \frac{e}{2\pi}\left(\frac{1}{\kappa_{\text{inc}}^{(2)}} - \frac{1}{\kappa_{\text{inc}}^{(1)}}\right)\right|^2 \frac{\left(k_x^2+k_y^2\right)k_z^{(1)}}{\omega\varepsilon_0}, \quad \text{(S18)}$$

which is symmetric with respect to $k_x$ and $k_y$, therefore the momentum-resolved cathodoluminescence spectrum is decomposed as

$$\Gamma^{\text{CL}}\left(k_{\|},\omega\right) = \Gamma^{\text{TR}}\left(k_{\|},\omega\right) + \Gamma^{\text{FP}}\left(k_{\|},\omega\right) + \Gamma^{\text{INT}}\left(k_{\|},\omega\right), \quad \text{(S19)}$$

with the individual components given as

$$\Gamma^{\text{TR}}\left(k_{\|},\omega\right) = \frac{e^2}{4\pi^2}\left|\frac{1}{\kappa_{\text{inc}}^{(2)}} - \frac{1}{\kappa_{\text{inc}}^{(1)}}\right|^2 \frac{k_{\|}^2 k_z^{(1)}}{\omega\varepsilon_0}, \quad \text{(S20a)}$$

$$\Gamma^{\text{FP}}\left(k_{\|},\omega\right) = \left| A_1^z + A_2^z e^{+ik_z^{(2)}d}\right|^2 \frac{k_{\|}^2 k_z^{(1)}}{\omega\varepsilon_0}, \quad \text{(S20b)}$$

and

$$\Gamma^{\text{INT}}\left(k_{\|},\omega\right) = \frac{e}{\pi}\text{Re}\left\{\left(A_1^z + A_2^z e^{+ik_z^{(2)}d}\right)\left(\frac{1}{\kappa_{\text{inc}}^{(2)}} - \frac{1}{\kappa_{\text{inc}}^{(1)}}\right)^*\right\}\frac{k_{\|}^2 k_z^{(1)}}{\omega\varepsilon_0}, \quad \text{(S20c)}$$

where $k_{\|}^2 = k_x^2 + k_y^2$. Hence, the total CL spectrum is decomposed in 3 terms, namely, the transition radiation contribution from the top surface ($\Gamma^{\text{TR}}\left(k_{\|},\omega\right)$), the Fabry–Pérot resonances ($\Gamma^{\text{FP}}\left(k_{\|},\omega\right)$), and the interference term ($\Gamma^{\text{INT}}\left(k_{\|},\omega\right)$).

In deriving the cathodoluminescence spectrum, we have neglected the power associated with the electron's self-field, i.e., the power generated by the electron current itself, as well as the interference terms between the electron's self-field and the scattered-field contributions, as the field accompanying a monotonically moving electron does not contribute to the radiation continuum.

The transition radiation from the top surface does not depend on the scattered field amplitudes ( $A_{1,2}$ as well as $A_{r,t}$ ) and it remains purely dependent on the materials specifications (RPP's permittivity) and the electron velocity. Therefore, regardless of the thickness of the RPP layer, the transition radiation contributions demonstrate the same intensity and spectral features (Supplementary Fig. S3d).

It should be mentioned that generally the adopted understanding of the transition radiation mechanism considers the summations of $\Gamma^{\mathrm{TR}}\left(k_{\|}, \omega\right)+\Gamma^{\mathrm{FP}}\left(k_{\|}, \omega\right)$, which directly include coherent modal excitations as well as the transition radiation from the top surface[6,7]. Here, we decompose these two terms, since the first term occurs naturally in all electron energies that traverses the surface, regardless of whether the electron can reach the lower surface or not. This decomposition has allowed us to configure the CL responses caused by the different electron energies.

In contrast to the transition radiation, the contributions of the Fabry–Pérot resonances within the RPP layer to the CL spectrum demonstrates a clear thickness-dependence, and also constitutes the dominant signal, as understood by comparing the decomposed CL intensities (Supplementary Fig. S3d to f). Comparing the total CL intensity (Supplementary Fig. S3g) to the eigen modes of the system depicts the ability of the CL to manifest the Fabry–Perot resonances within the light cone, whereas the signature of the guided wave modes with their dispersions positioned outside the light cone is captured near to the light cone as well. The modal dispersion is calculated by considering the zeros of the determinant shown in Supplementary Eq. S17a. The plot presented in Supplementary Fig. S3c is $\left|\Delta\left(k_{\|}, \omega\right)\right|^{-1}$.

To better understand the role of the gold substrate in enhancing the Fabry–Pérot resonances, we compare the modal dispersion retrieved as $\left|\Delta\left(k_{\|}, \omega\right)\right|^{-1}$ for the cases with and without the gold substrate (Supplementary Fig. S4a and b). The inclusion of the gold substrate leads to a blue shift of the Fabry–Perot resonances, and the appearance of surface plasmon resonances at the interface between the RPP and gold, with the latter demonstrating the strongest contribution; however, its dispersion is positioned outside the light cone and is not directly accessible through our CL technique. The presence of gold substrate results as well on connecting the Fabry–Perot and guided-mode resonances, with localized resonances appearing at the light line and becoming accessible with the CL technique, when the numerical aperture of the collecting optic approaches unity.

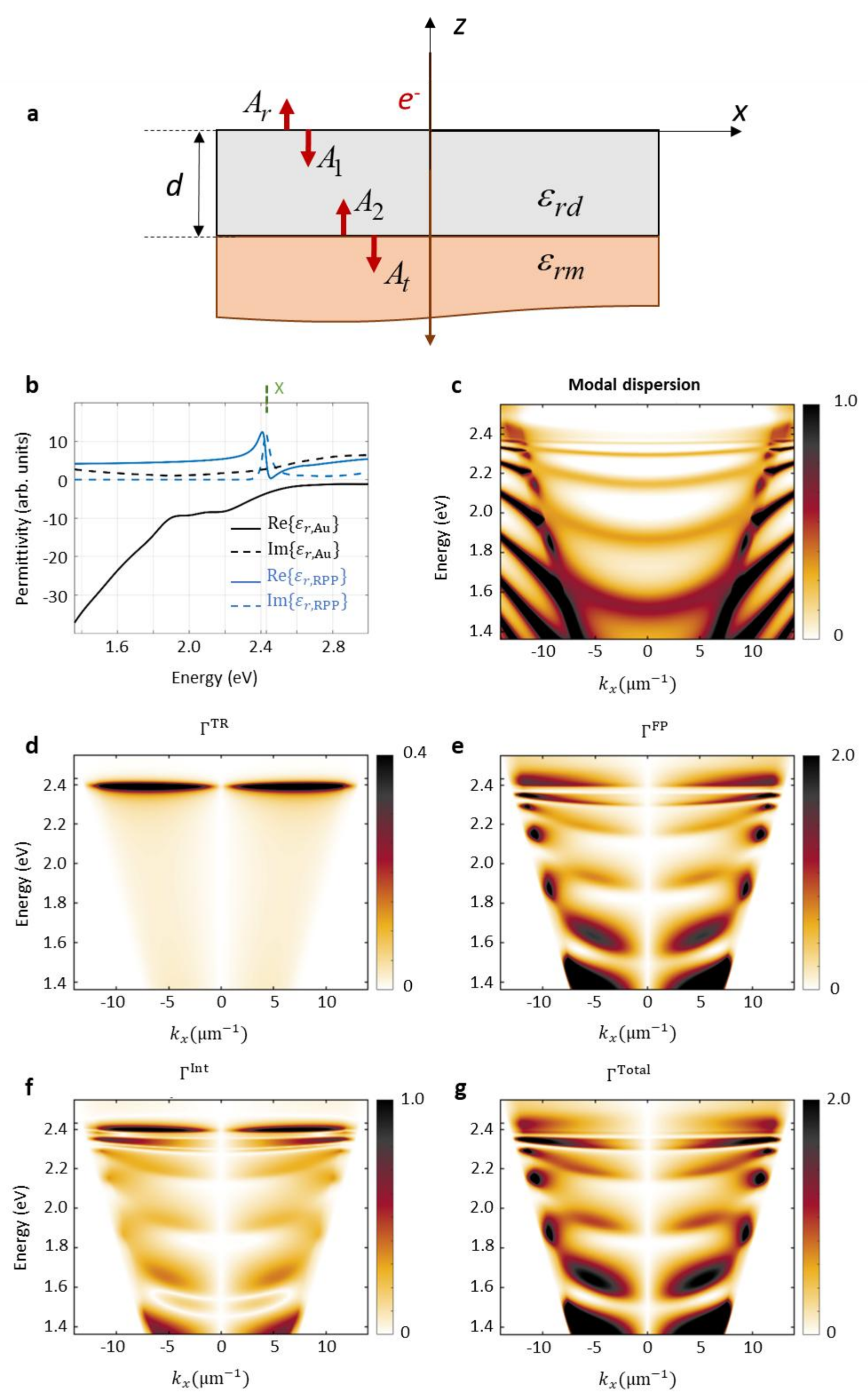

**Supplementary Figure S3. Composition of the cathodoluminescence radiation.** (**a**) Relative permittivity of $BA_2PbI_4$ RPP (blue lines) and of gold (black lines), the latter is used as substrate. The real and imaginary parts of the permittivity are depicted by solid and dashed lines, respectively. (**b**) Calculated modal dispersion of a 680 nm thick RPP layer on gold, showcasing Fabry–Pérot modes inside the light cone and waveguide modes outside of it, both bent due to a strong coupling to the exciton resonance. (**c-f**) Calculated radiation components under normal-incidence electron beam excitation. Here, (c) shows the transition radiation, which peaks at the energy associated with the maximum of the real part of RPP permittivity. The polaritonic cavity modes of the structure coupling to light cause the signal depicted in (d). Together with the interference between the transition radiation and Fabry–Perot resonances (e), the sum of all radiation contribution from the total signal displayed in (f).

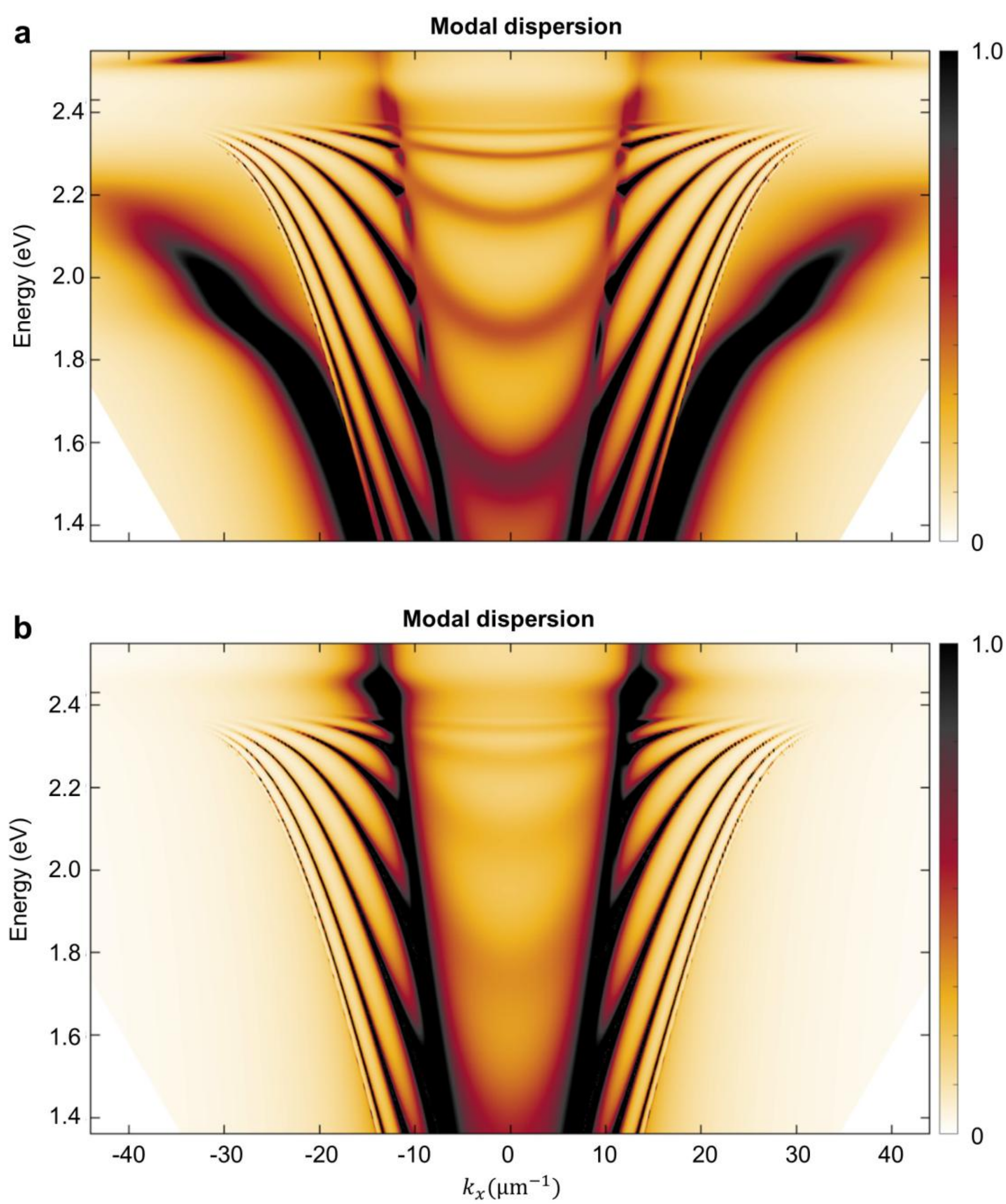

**Supplementary Figure S4. Substrate dependence of the modal dispersion in RPP layers.** (**a**) Analytically calculated modal dispersion exhibited by the air-RPP-gold layered system, yielding polaritonic Fabry–Pérot modes inside of the light cone and waveguided polaritonic modes between the vacuum and the RPP light line, as well as a transition area at the light line due to the half-open cavity nature of the system. The broad mode beyond the RPP light line is due to the surface plasmon polariton excited at the RPP-gold interface. (**b**) Modal dispersion of the symmetric air-RPP-air system as a comparison, showing a clear separation of waveguided and Fabry–Pérot modes at the vacuum light line and decreased intensity of the latter due the reduced cavity quality.

**Supplementary Note S3: Photoluminescence excitation spectroscopy**

Due to the large material dissipation in RPP above the exciton energy, upper polariton branches (UPBs) are not directly distinguished in the cathodoluminescence and the photoluminescence spectra. Nevertheless, they can be identified using Photoluminescence excitation spectroscopy. When resonantly excited with laser, the population in the UPBs can decay to the LPBs. Hence, they can be probed by tuning the laser excitation wavelength measuring the corresponding photoluminescence efficiency through LPB channels. By exciting the upper polariton branches directly, the measured populations of LPBs are maximized and subsequent effects on the polaritonic spectra can be investigated. The integrated photoluminescence spectra at different excitation wavelengths for a thick RPP flake on gold is shown in Supplementary Fig. S5c for the flake shown in Supplementary Fig. S5a, where the spectra were normalized to the wavelength-dependent laser intensity. A peak in excitation efficiency is observed at around 460 nm, showing an upper polaritonic branch, or the peak of a superposition of multiple upper polaritonic branches.

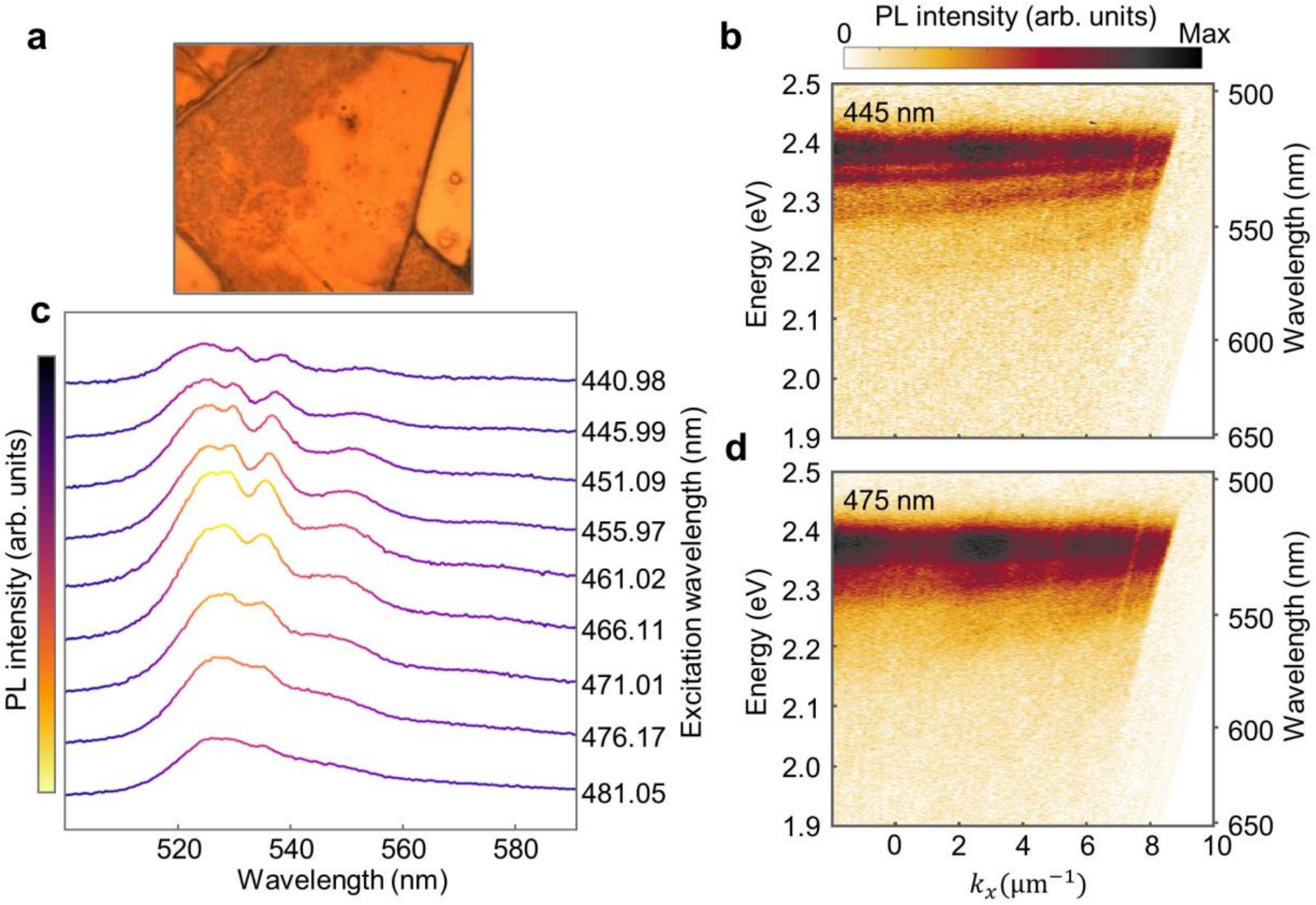

**Supplementary Figure S5. Polaritons in photoluminescence excitation spectroscopy of RPP on gold. a**, Reflection image of the investigated RPP flake on gold. **b**, Angle-resolved photoluminescence spectrum at 445 nm laser excitation, displaying a multitude of excited lower polariton modes. **c**, Excitation wavelength scan with spectra averaged over $k_x$ from $k_x = -2\ \mu m^{-1}$ to $k_x = 2\ \mu m^{-1}$. The spectra exhibit a peak excitation efficiency around 460 nm likely related to upper polariton excitation and show degradation effects at higher wavelengths since the scan started at 441 nm. **c**, Angle-resolved photoluminescence spectrum at 475 nm laser excitation with a clear broadening of the polaritonic modes due to degradation.

Additionally, a blue shift of the lower polaritonic branches with increasing wavelength is visible, but in this case, it has to be taken into account that all measurements were performed on the same spot. Thus, because the scan started at low wavelengths, the shift is likely degradation-dependent, and a similar effect is evident in a degradation test measurement depicted in Supplementary Fig. S7. The degradation becomes obvious when comparing the angle-resolved spectrum at 445 nm with that of 475 nm excitation wavelength. With a severe degradation, the lower polaritonic branches broaden and overlap, indicating a reduced cavity quality.

### Supplementary Note S4: Effect of excitation intensity and degradation in photoluminescence spectroscopy

We have conducted intensity-dependent photoluminescence measurements on a thick RPP flake positioned on gold by employing neutral density filters. The corresponding angle-resolved maps are shown in Supplementary Fig. 6a with a filter with an optical density (OD) of 2, in Supplementary Fig. 6b with OD 1, and in Supplementary Fig. 6c without a filter, thereby increasing the intensity roughly by 10 times with each step. No obvious change in level repulsion is found, and additionally, only a small shift in population towards lower polaritonic branches takes place with increasing intensity. This yields the conclusion that within the intensity ranges of our system no clear polariton-polariton interaction—and as expected —no collective coherent exciton-photon interaction is evident.

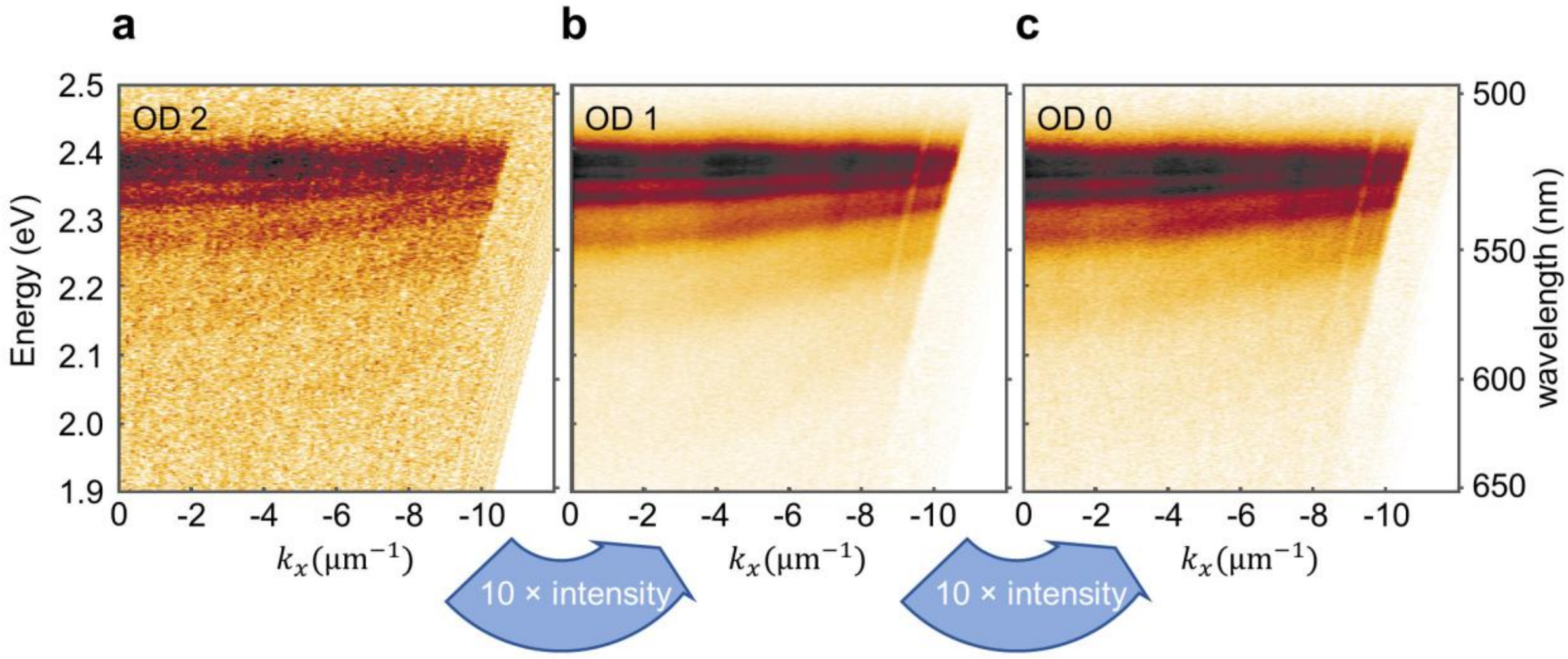

**Supplementary Figure S6. Intensity-dependent optical excitation of polaritons.** (**a**, **b**, **c**) Angle-resolved photoluminescence spectra of a thick RPP flake on gold at 476 nm excitation wavelength, where the laser is filtered by a neutral density filter with an optical density (OD) of 2 **(a)**, of 1 **(b)** and of 0 **(c)**. The images are normalized and integration times vary to avoid sample degradation, leading to changing signal-to-noise ratios. No obvious mode shift is observed in the power ranges of our laser system.

If now the laser intensity is kept constant and the spectra are recording after various illumination times, the photo-induced degradation effect can be analyzed. Supplementary Figure 7a and b show a polaritonic spectrum after 5.3 s and after 90.5 s illumination time with the same acquisition time. Clearly, a decrease in photoluminescence intensity is caused by the degradation. Moreover, by integrating over a small range around the in-plane wave number around $k_{||} = 0\ \mu m^{-1}$, the degradation-related change in the spectral shape is visualized (Supplementary Fig. S7c). A broadening and blue shift of the lower polaritonic modes is observed, with the former indicating reduced cavity quality due to a surface roughening effect. Since Fabry–Pérot modes are blueshifted in thinner cavities, this is a hint of photo-induced degradation thinning down the cavity, however, a more detailed investigation is beyond the scope of this work.

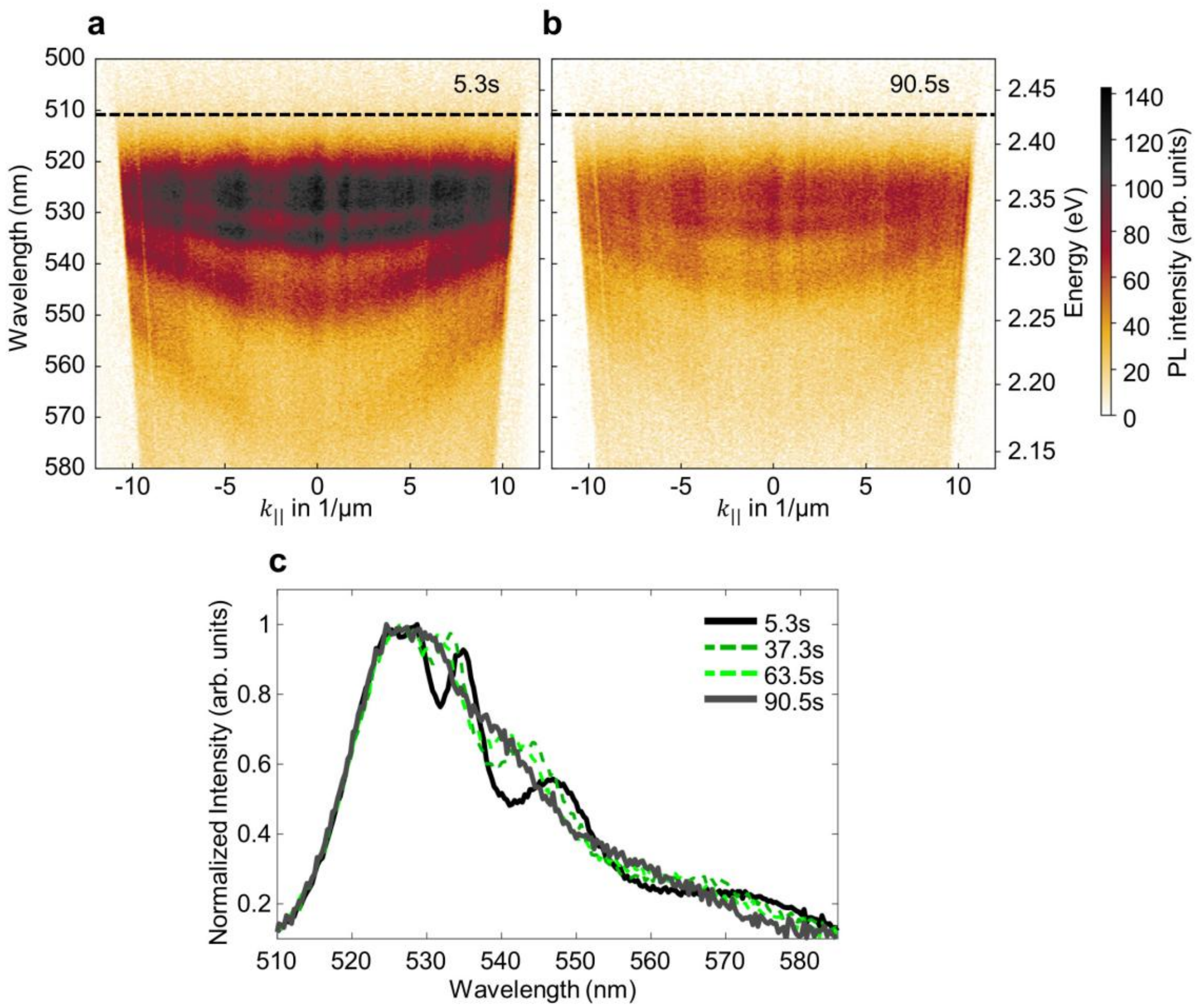

**Supplementary Figure S7. Photoinduced degradation effect on polaritonic dispersion relations in RPP.** Angle-resolved photoluminescence spectrum of a thick RPP flake on gold at 475 nm excitation wavelength **(a)** after 5.3 s and (**b)** after 90.5 s illumination time with constant laser power and the same acquisition time. (**c**) Normalized spectra integrated over a small range around the gamma point ($k_x = -2\mu\mathrm{m}^{-1}$ to $k_x = 2\mu\mathrm{m}^{-1}$) to illustrate the change in peak shape and position after different illumination times. Photoinduced degradation leads to a decrease in intensity, a broadening of the polaritonic modes and an apparent blue shift of the distinguishable polaritonic modes.